# Switched Mode Four-Quadrant Power Converters

*Y. Thurel*
CERN, Geneva, Switzerland

**Abstract**
This paper was originally presented at CAS-2004, and was slightly modified for CAS-2014. It presents a review of the key parameters that impact the design choices for a true four-quadrant power converter, in the range 1–10 kW, mainly based on switching mode converter topology. This paper will first describe the state-of-the-art for this power converter family, giving the drawbacks and advantages of different possible solutions. It will also present practical results obtained from the CERN-designed converter. It will finally give some important tips regarding critical phases like test one, when conducting a project dealing with this type of power converter.

## 1 Introduction

The design of a four-quadrant power converter design is strongly dependent on its use, and several criteria are reviewed in the first part of this document. A review is then made of some commonly used topologies. The second part of this document gives the key points of the design for a specific four-quadrant power converter designed for powering superconductive magnets in the LHC machine.

In this paper the load will always be assumed to be a magnet, with resistance and inductance, which is a realistic four-quadrant power load. Since the system studied is highly non-linear and some topologies are quite complex to simulate (with up to three control loops working at the same time), the author gives a lot of illustrated key points instead of formulas. This approach is easier to follow as known problems are solved during the design phase. This paper mainly focuses on low or medium output voltages in the range −200 V to 200 V. For a high voltage converter, a different approach would normally be required, since the available components (semiconductor or passive) will have a big impact on the final design choice.

## 2 Definitions and description of different quadrant operation

The definition and the numeration of operating quadrants are shown in Fig. 1. The naming of each *generator* and *receptor* quadrant has been done according to the energy being managed by the converter placed between the energy source (mains) and the load. Energy can be taken from the mains and delivered to the load, the converter voltage and current being of the same sense, with the converter being seen to *generate* energy from the load's point of view. In the case where the voltage and current are opposite, energy comes back from the load, being 'received' by the converter managing it, in *receptor* mode.

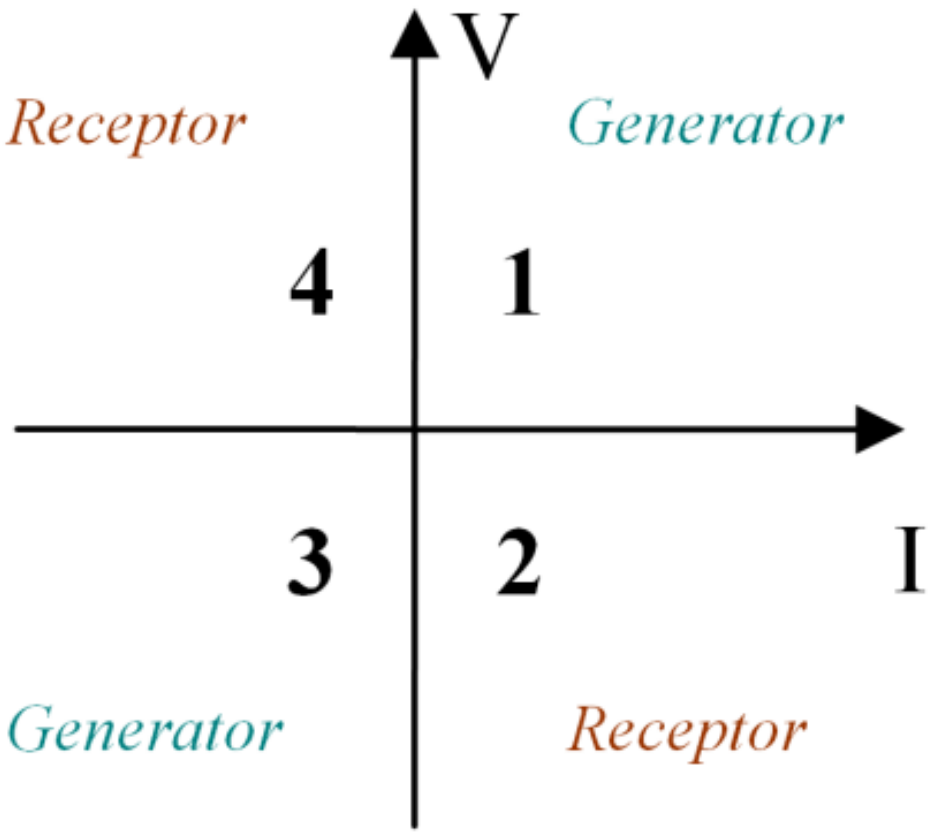

**Fig. 1:** Typical one-quadrant converter naming conventions

Some simplified graphs are presented to summarize the different type of power converter, applied to a typical four-quadrant power load, with a resistance in series with an inductance (R-L). Through this load example, the limitations of different topologies can be illustrated.

Figure 2 shows a possible curve for a one-quadrant converter, where the current can be controlled while it is increasing. The control of a decreasing current is still possible, but drastic conditions and limitations must be taken into consideration at the level of the power system definition; and also at the level of the current controller. If a negative ramp applied to the current is faster than the load's natural constant time, control will be lost. Also, a commercial one-quadrant converter can sometimes exhibit high output capacitance (for example in the case of a laboratory power supply), with a non-symmetrical rate of voltage change across the capacitor related to the load's operating conditions, i.e. mainly its level of current.

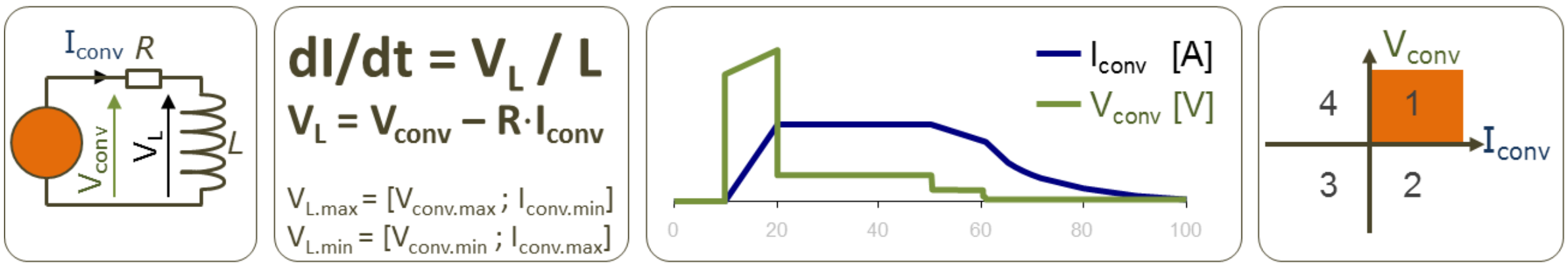

**Fig. 2:** One-quadrant converter typical curves

Figure 3 shows a possible curve for a two-quadrant converter, where the current can be controlled as long as it stays positive-only (or negative only). While the current decreases, and depending on its required ramp rate, the converter can be required to absorb energy from the load. In such a case, load energy will be 'removed' from it, being dissipated in cable resistance, but also 'managed' by the converter if its extraction is required to be faster than the rate allowed by the resistance. The converter can or dissipate, store, or send back the received energy to the original energy source (the mains). It should be understood that the converter must control the level of energy flowing from the energy source, through the converter, to the load inductance, or back from the inductance, in a fully controlled way.

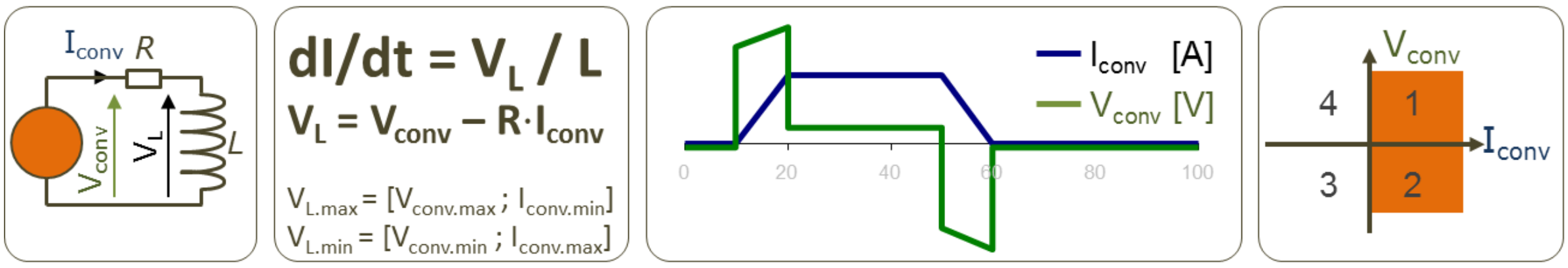

**Fig. 3:** Two-quadrant converter typical curves

There should be no confusion with some converters being operating in two diagonal quadrants, for example quadrants one and three, by the addition of a polarity switch (mechanical or electronic inverter); the converter, often a one-quadrant converter, will be able to deliver positive and negative current to the load, but will not be able to recover any energy from the load. Despite the load current being positive or negative, this kind of converter will suffer from exactly the same limitations as a one-quadrant converter, regarding the possible current ramp and controllability.

Figure 4 shows a possible curve for a four-quadrant converter, where the current can be controlled without any limitations, regarding to its sense, or evolution (positive or negative ramp). Like a two-quadrant-power converter, the converter can be required to absorb energy from the load. If some topologies are intrinsically providing full and true four-quadrant operation capability like an H-bridge, the management of energy and the directionality of the output current can often lead to relatively complex designs.

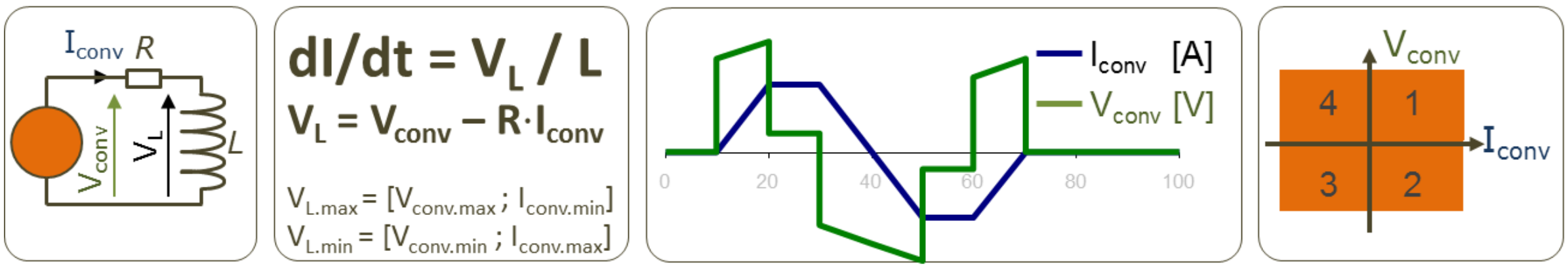

**Fig. 4:** Four-quadrant converter typical curves

## 2.1 Receptor mode solutions, a brief review

In the receiving quadrants (quadrants two and four), the converter has to extract energy from the load, either dissipating it, storing it, or sending it back to the mains. It should be remembered that a true four-quadrant power converter should regulate its output conditions while 'absorbing' energy, not being limited in its controllability and performance due to the quadrants in which it operates.

### 2.1.1 *Locally storing the load energy*

This function is often performed using capacitors, which are in charge of collecting load energy; these capacitors can be a naturally part of the topology chosen (full-bridge DC–DC topology), and dimensioned to take into account this secondary function (storage capacity). This storage capability can be dedicated either on the primary or secondary sides; this choice is made depending on the voltage used at the output, and also taking into consideration the complexity of the final design.

Even if some designs already include the local storage of energy, supercapacitors or superconducting inductors can present an attractive alternative in the case of demanding energy management, pushing this topology ahead. Of course, a variety of possible 'mechanical' storage solutions also exists, like kinetic energy storage, which relies on a rotating machine being coupled to the existing topology; these solutions, which are out of the scope of four-quadrant switching converters, are not treated in this paper.

### 2.1.2 *Sending the load energy back to the mains*

This field is dominated by thyristor-based topologies (two thyristor converters operating back-to-back), and is still extensively used when high power is required. Simplicity of design and robustness of this well-known topology are other big advantages. The price to be paid is certainly the limited bandwidth to be expected from network based on natural low frequency Also, some high frequency switching power converter designs have been presented as a valid alternative, with a working prototype validated at CERN, with relatively complex control requirements [2].

### 2.1.3 *Dissipating the load energy*

This is surely the least appealing solution, especially when semiconductors are used to dissipate energy as pure heat losses. Nevertheless, this alternative solution can be integrated into modern topologies, and still presents good performance in the fields of electromagnetic compatibility (EMC) and very high bandwidth performance. This approach can be justified when performance is judged to be the most important consideration, with the frequency of charge/discharge that the load will require determining viability at a power management level.

## 2.2 Influence of load cycle on converter topology

Different types of accelerators use different types of magnet for different purposes. It could be drastically summarized in two main different domains: pulsed or slow, as represented in Fig. 5. When a pulsed machine is designed the energy-saving aspects should be taken into account. With high frequency charge and discharge sequences, a true four-quadrant converter, giving back energy to the mains, or locally storing it for the next run, should be envisaged. In the other hand, a slow and ultra-high precision machine (like the LHC) will focus on different criteria like stable controllability versus operating point and the low-level conducted noise environment (EMC), which are both crucial criteria for high precision operation.

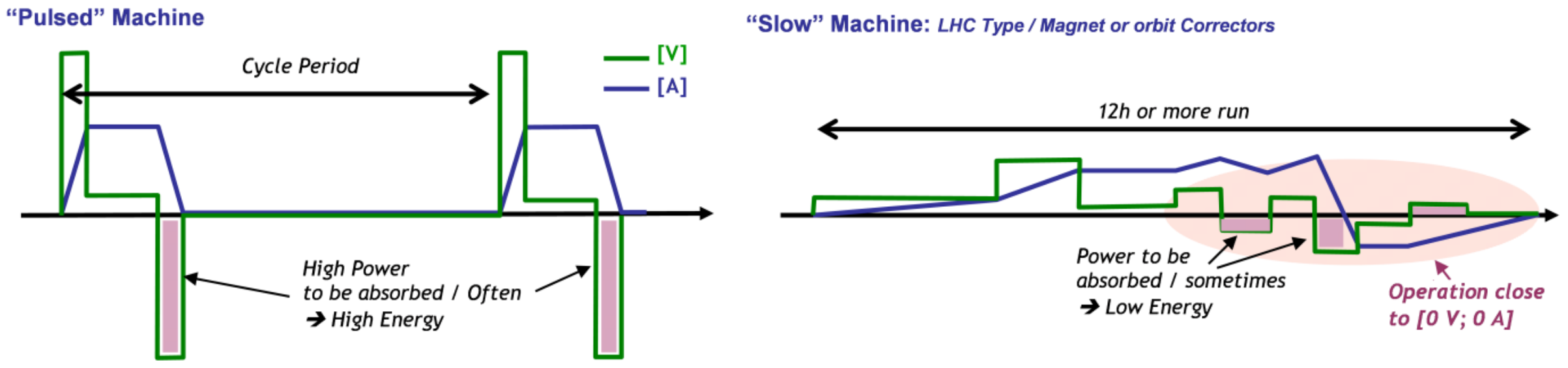

(a) (b)

**Fig. 5:** Machine type operation typical cycles. (a) Pulsed machine; (b) 'slow' machine

## 2.3 Load influence on design parameters

The LHC requires a very high magnetic field to control the beam; this field, being directly proportional to the current, can require up to 600 A load current in some four-quadrant converter families. The number of correctors, combined with these levels of current, made the superconductor magnet a valid choice for energy-saving considerations. These magnets are by definition lossless loads, since magnet is a pure inductance, up to some Henrys. To avoid adding extra losses in the tunnel, where converters are installed close to their loads, large cross-section copper cables were connected to these magnets, connecting the power converter to its load. All of these boundary conditions lead to a very high time constant circuit (large inductance value combined with very low resistance value), keeping in mind that:

$$\tau_{CIRCUIT} = \frac{L_{MAGNET}}{R_{CABLE}} \quad , \tag{1}$$

where $L_{\mathrm{MAGNET}}$ is the magnet inductance value [H] and $R_{\mathrm{CABLE}}$ is the cable resistance value [Ω].

The value of this time constant, inherent from the circuit type and physical characteristics, combined with the LHC's required characteristics giving the range of the current, determines the type of power converters to be used with specific operating areas,

$$P(I) = U(I) \cdot I = R_{\mathrm{CABLE}} \cdot I^2 + L_{\mathrm{MAGNET}} \cdot \frac{\mathrm{d}(I)}{\mathrm{d}t} \cdot I \; . \tag{2}$$

It can be stated that a trade-off could be found, given an operating range ($I_{MAX}$ and d$I$/d$t$) for a dedicated magnet ($L_{MAGNET}$ fixed) between generating peak power and regenerating absorbed peak power, modulating the cable resistance. Of course, if increasing cable resistance mechanically increases the losses, it will nevertheless decrease power absorbed by the power converter. Increasing regenerating power for a four-quadrant converter can be a lot more difficult than simply slightly increasing its generating power, saving a lot on the regenerating power level.

Figure 6 shows the typical case of a given superconducting magnet used with two different resistances given by the cable, 10 mΩ and, doubling the cable resistance, 20 mΩ.

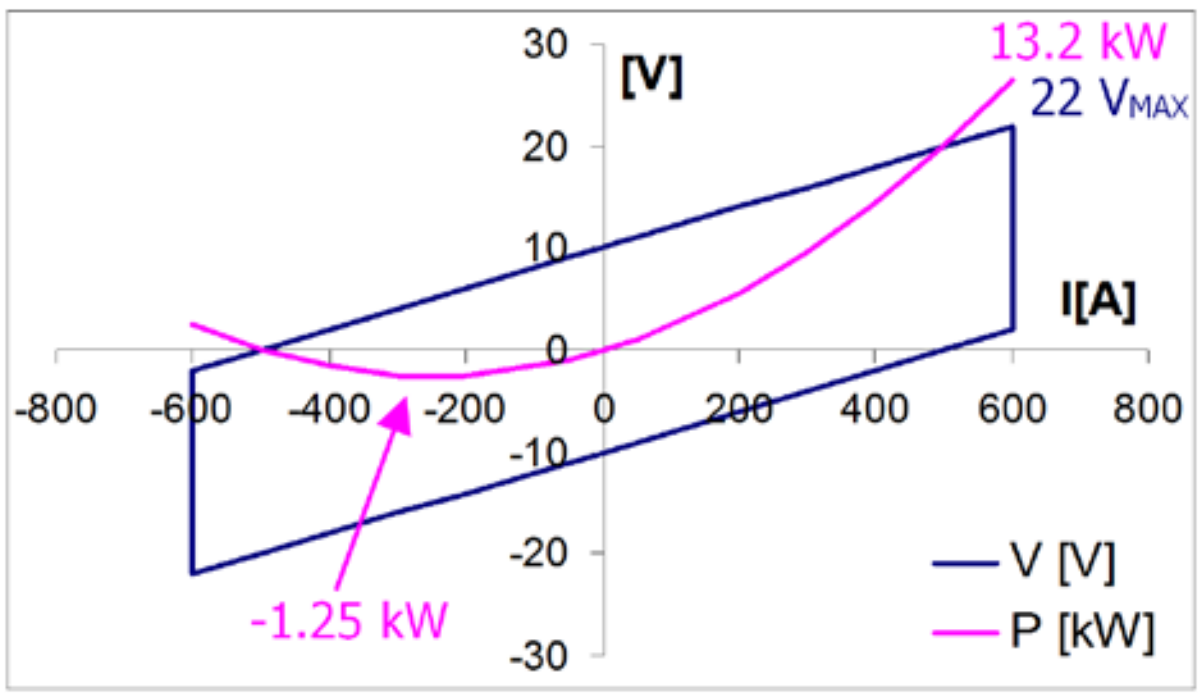

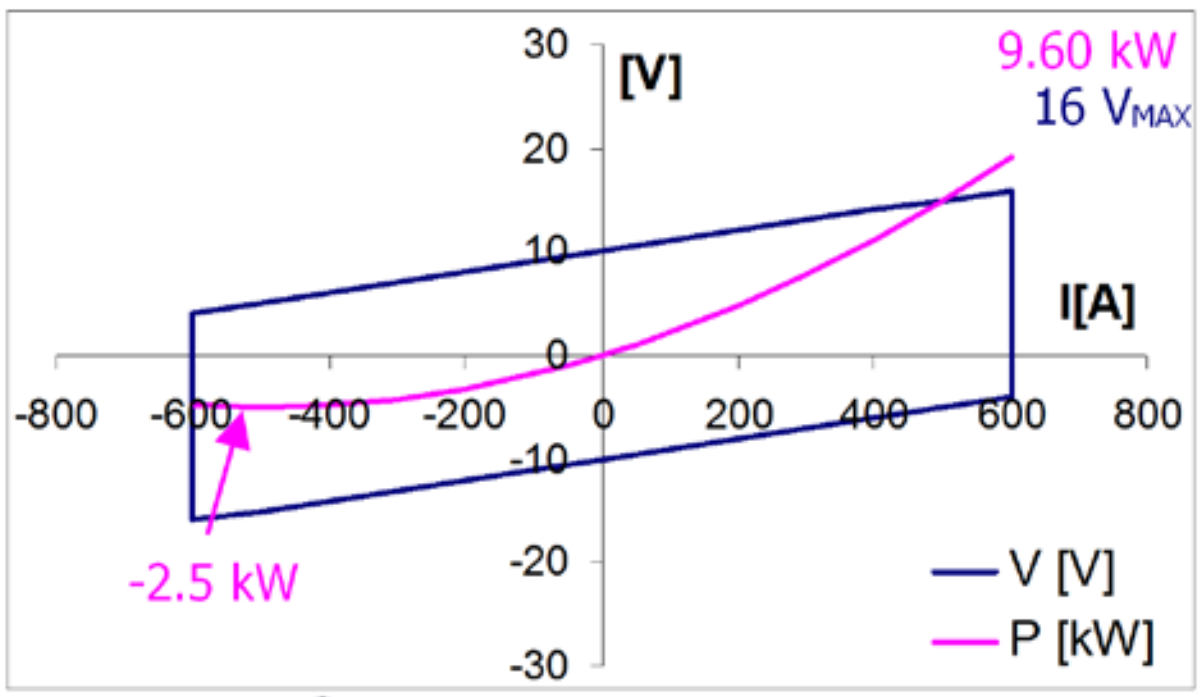

(a) (b)

**Fig. 6:** Influence of circuit parameters on converter operation. (a) 20 mΩ; (b) 10 mΩ

In the 20 mΩ case described above (Fig. 6(a)), the power to be generated is 1.4 times higher when the level of power to be absorbed is up to two times lower, and at a lower operating current. Of course, cable losses are two times higher during all magnet operations. Nevertheless, this consideration could be really interesting in the case of a corrector magnet, which is not often used at full current (less heat losses), lowering the power converter design constraints.

# 3 Main topologies for the four-quadrant stage

A four-quadrant stage is the dedicated part of a power converter used to manage load voltage and current in the four-quadrant area. This function can be part of the converter topology (thyristor-based or H-bridge DC–DC topologies) or as a kind of extension to a standard one-quadrant power converter. This section deals with classical solutions, which are mainly used in the four-quadrant power converter domain.

## 3.1 Two thyristor bridges mounted in anti-parallel

This standard solution makes it possible to send back energy to the mains. It is based on two thyristor bridges, mounted in anti-parallel using the natural two-quadrant capability of each bridge. In this case, the four-quadrant stage is represented by the whole thyristor converter for half of the quadrant plane. This solution is very well known and can handle high power constraints. If noted here as a reminder, this topology will not be described in detail since it is not usually part of the switched mode converter family.

## 3.2 Linear dissipative stage

### 3.2.1 *Description*

A linear dissipative stage relies on a push–pull stage, with transistors used as ‘programmable resistors’ dissipating energy in their receiving modes. This stage is usually an additional stage for a standard generator power converter that is used to provide power at the input of the linear stage.

An alternative solution exists with polarity switches added to transform a single output to the required double outputs. This solution, see Fig. 7, is much easier considering the power converter to be designed, since it has a single output, but it leads to potential distortions that are seen when the polarity switches act. It also requires the control of additional transistors, while the standard double output voltage source can use simple additional rectifiers (an L-C filter can be shared between the two outputs, reducing the number of additional components and cost).

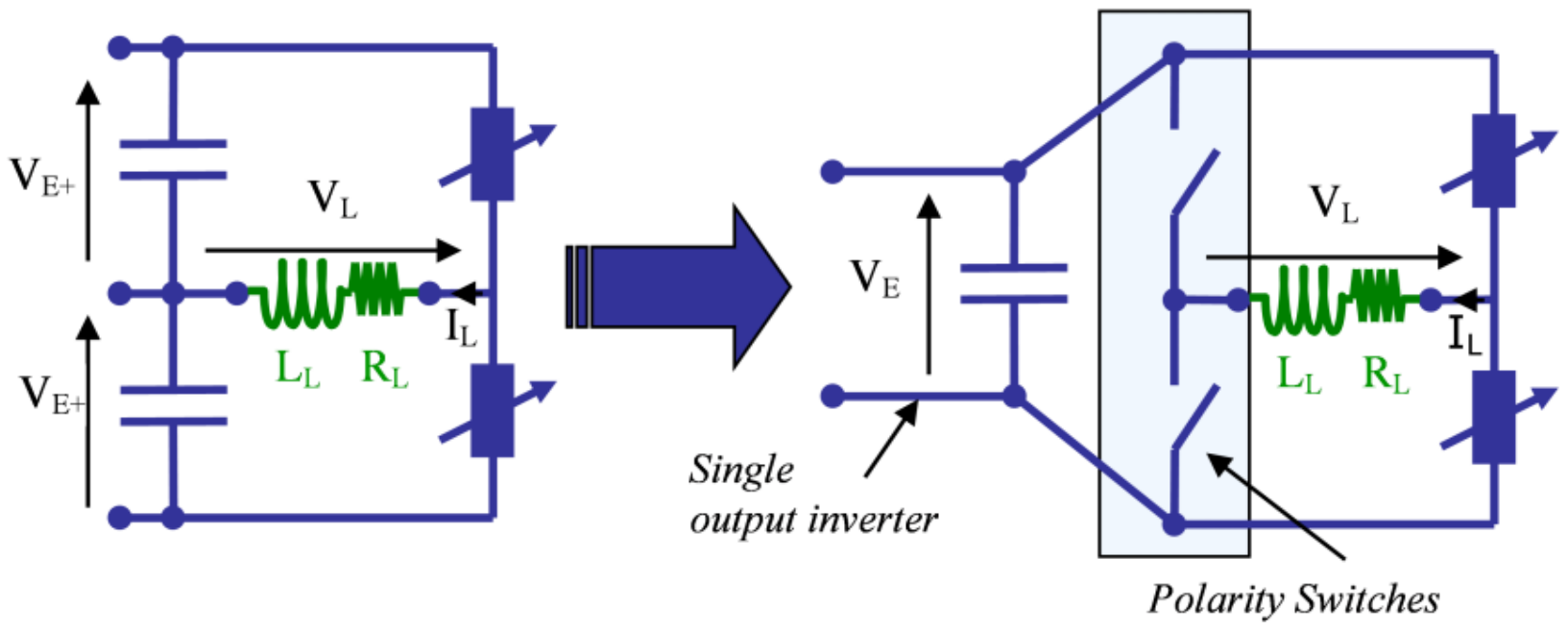

**Fig. 7:** Linear stage schematic

### 3.2.2 *Operation principle*

A dual output power source is used to power the linear stage, with the usual limitation of the two dual outputs being tied together. (Standard solutions are usually based on a dual output power converter using one inverter stage powering a dual output transformer.) Two different situations are possible, with a fixed value or variable dual output configuration.

### 3.2.3 *DC fixed dual output configuration*

Figure 8 shows voltage and current waveforms for a typical magnet application, highlighting the limitations and dissipation problems to be dealt with. In the particular example below, a [±10 V; ±120 A] converter is shown. Since losses in transistors are assumed to be 2 V at the highest level of current, a minimum dual voltage of 12 V is required to feed the output linear stage.

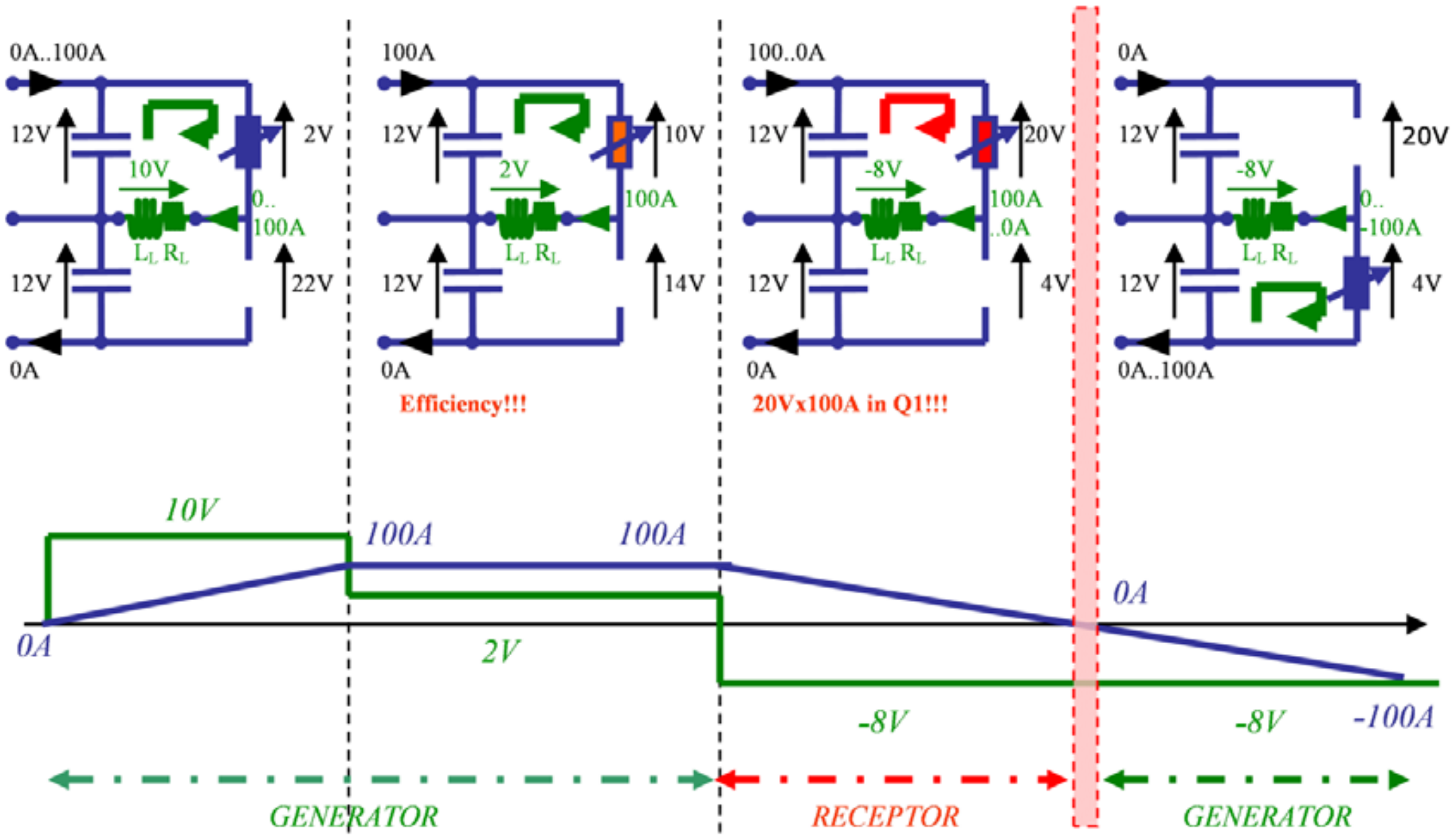

**Fig. 8:** Four-quadrant linear stage (fixed DC dual voltage)

A dual output 50 Hz transformer, with adequate rectifying and filtering series stage, can simply provide the two desired fixed DC sources, represented on the schematic by capacitor; modern design would potentially propose a switched-mode power supply for size reduction, for example. Both voltage sources have to provide a voltage capable of compensating the losses given by the voltage drop across the active switch at maximum current.

If this solution is very simple, an efficiency issue exists with loss management at the transistor level, especially in low voltage/high current conditions where the transistor has to dissipate a large proportion of the energy given by the power source. This is particularly true in superconducting magnet powering, since a high voltage is necessary to ramp the current to its maximum losses, when a steady state requires a very low voltage induced by the DC cable losses. A second source of losses is the regenerating phase, when an active transistor has to handle double the level of power that is normally required by the load.

An example of an old-fashioned system made of several transistors in parallel is shown in Fig. 9. If the system can be very fast (no topological power-side limitation except the transistor’s inherent speed), dissipation is a hot issue, as well as the potential for a cascading failure, which can result from an initial failure leading to the destruction of several transistors at the same time.

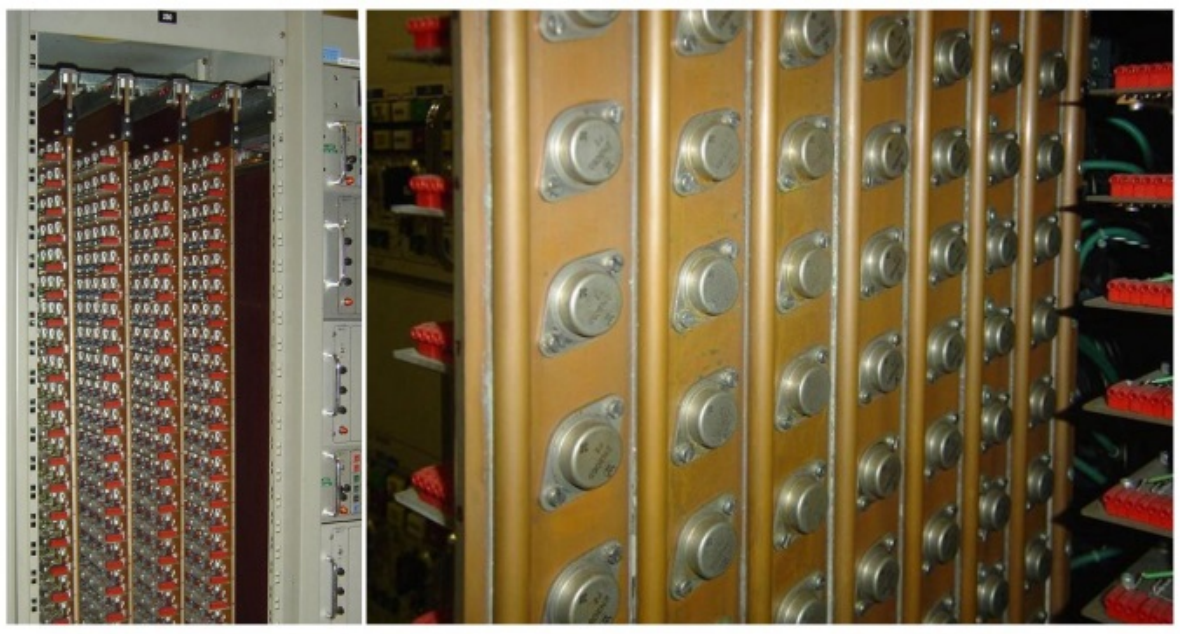

**Fig. 9:** Water-cooling linear stage of a CERN four-quadrant converter

### 3.2.4 *Variable DC dual output configuration*

The solution proposed in Fig. 10 is a lot better concerning the level of efficiency; generating power provided from the mains follows the requested load, in addition to the active transistor conduction losses.

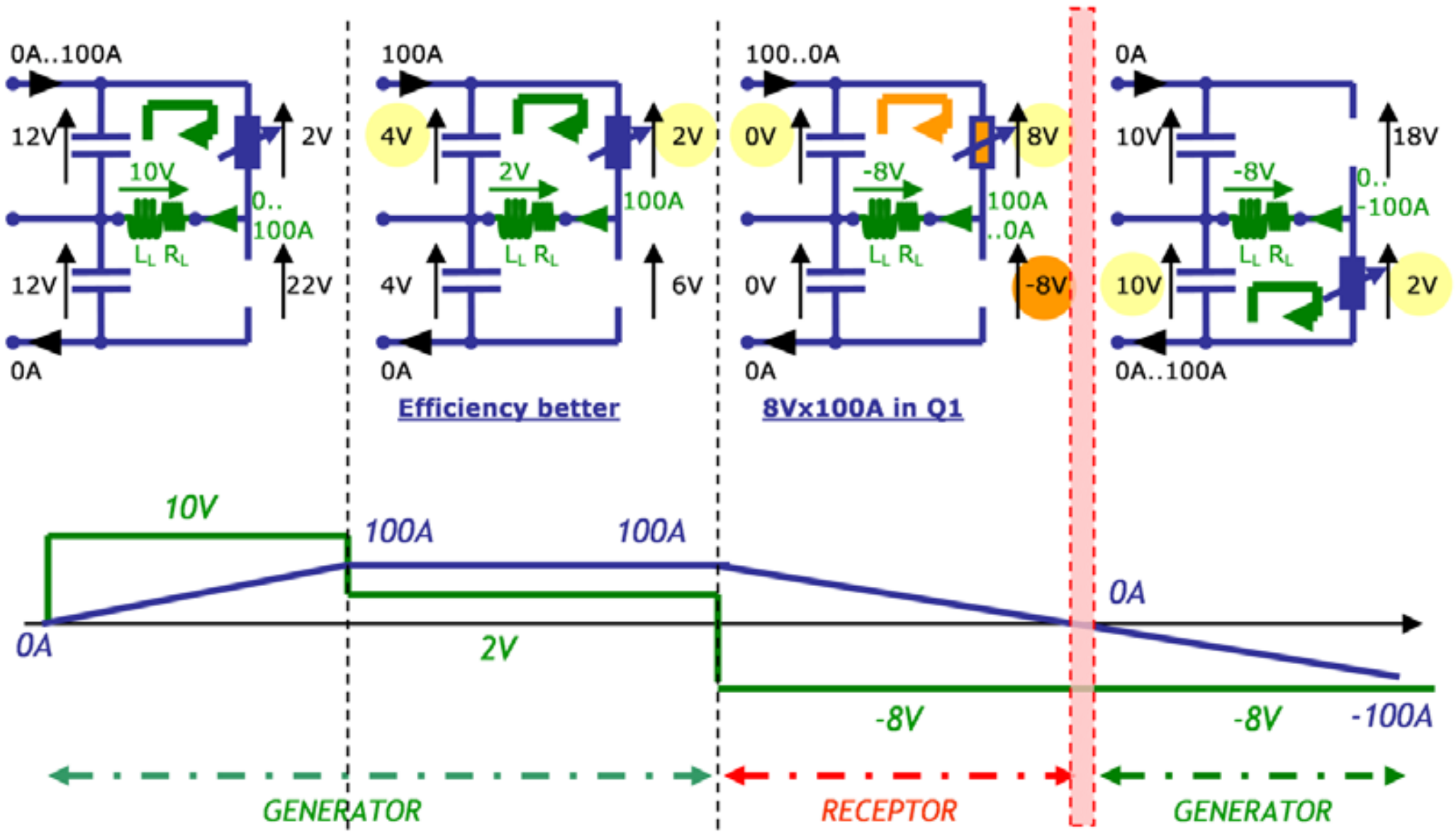

**Fig. 10:** Four-quadrant linear stage (variable DC dual voltage)

On the two schematics proposed, the 0 A zone is always considered to be a critical zone, since the linear stage always 'relies' on the current going through the relevant transistor following the voltage reference. This state is particularly critical when the 0 A point has to be crossed, while voltage has to be present on an inductive load. This is clearly impossible as shown in Fig. 10, and needs some artificial circuitry.

The voltage level across the transistor while the converter is operating in generator mode (2 V above) needs deep analysis. Indeed, it will be demonstrated below that it is possible to benefit from using an unsaturated linear stage, for giving extra rejection of mains perturbation. This option has obviously an efficiency cost since conducting active transistor losses will be higher.

## 3.3 Switching stage

### 3.3.1 *Description*

A conventional H-bridge stage (with an L-C filter in series to filter the switching ripple at the output level) can deal with energy sent to or received from the load. Energy has to be managed at the input of the H-bridge level, stored in the capacitor, or discharged by an additional brake chopper (switching transistors in series with a resistor, in charge of limiting the overvoltage condition across the input capacitor).

### 3.3.2 *Operation principle*

The H-bridge, see Fig. 11, is a natural four-quadrant power stage, which makes a good candidate for a four-quadrant power converter, combined with a single voltage power source, providing voltage adaption and insulation from the mains. This power source can be either a 50 Hz transformer or a modern switched-mode-based solution. This topology is very simple to control, since the duty cycle alone directly controls the output voltage, without any transition mode problem.

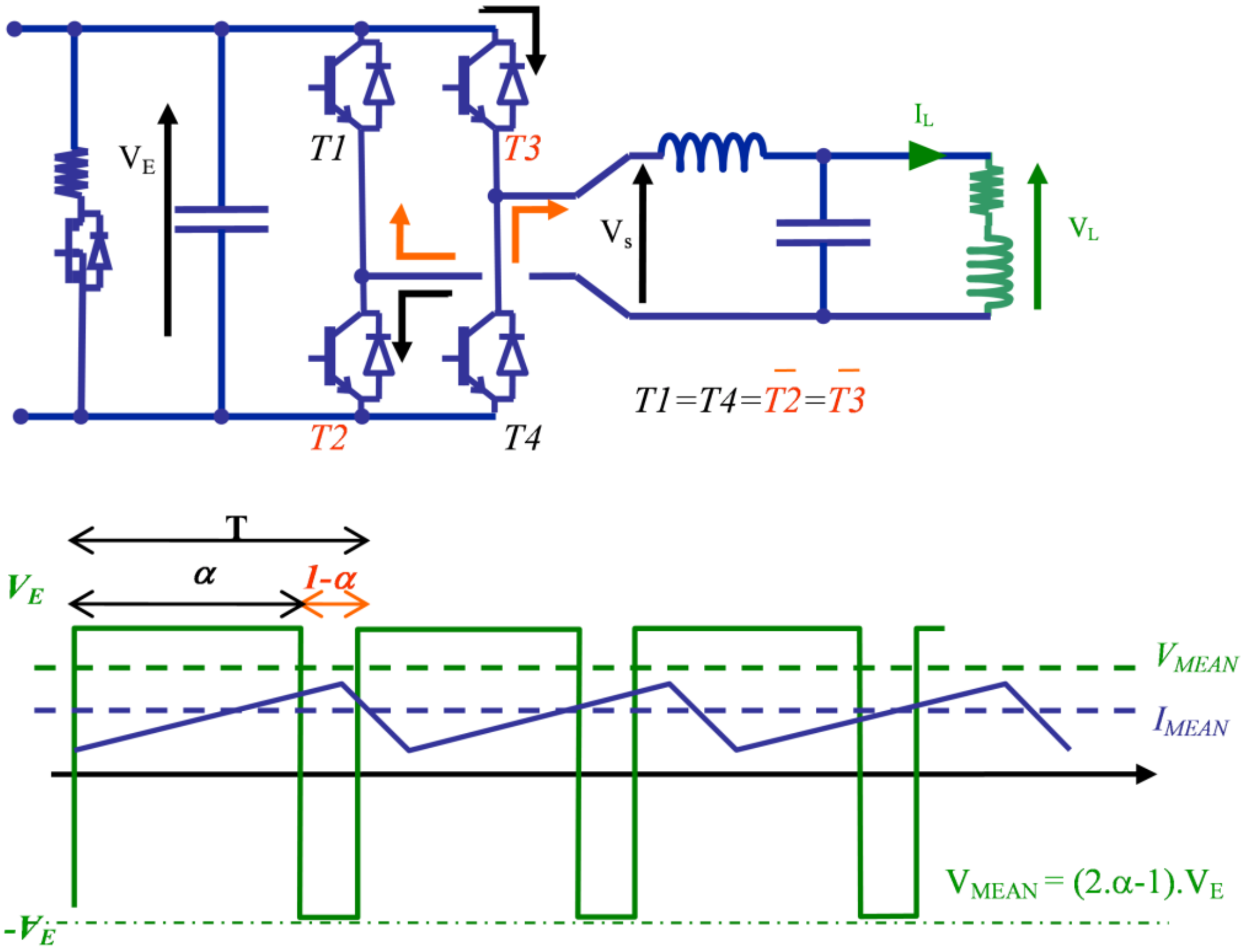

**Fig. 11:** Four-quadrant switching stage

This topology is often used, since it is highly flexible in terms of power, but it is also flexible in terms of voltage and current operating range. Concerns come from the losses due to the hard switching stage dealing directly with output current, and EMC at the output level resulting from the proximity with the switching cells. It must be taken into account that, once the load is grounded, the H-bridge capacitor, with the source it relies on, will be naturally excited at the switching frequency rate, with severe common mode noise issues to be solved at the output level.

In some advanced cases, converters can use a two-operation mode control, transforming the H-bridge into a classical buck converter, to reduce losses and current ripple (a leg is inactive when a switch is closed).

## 4 Review of different topologies

Figure 12 gives an overview of the possible combination of stages required to build a four-quadrant power converter. All ‘rectifier bridge’ paths indicate a topology where energy cannot be returned to the mains.

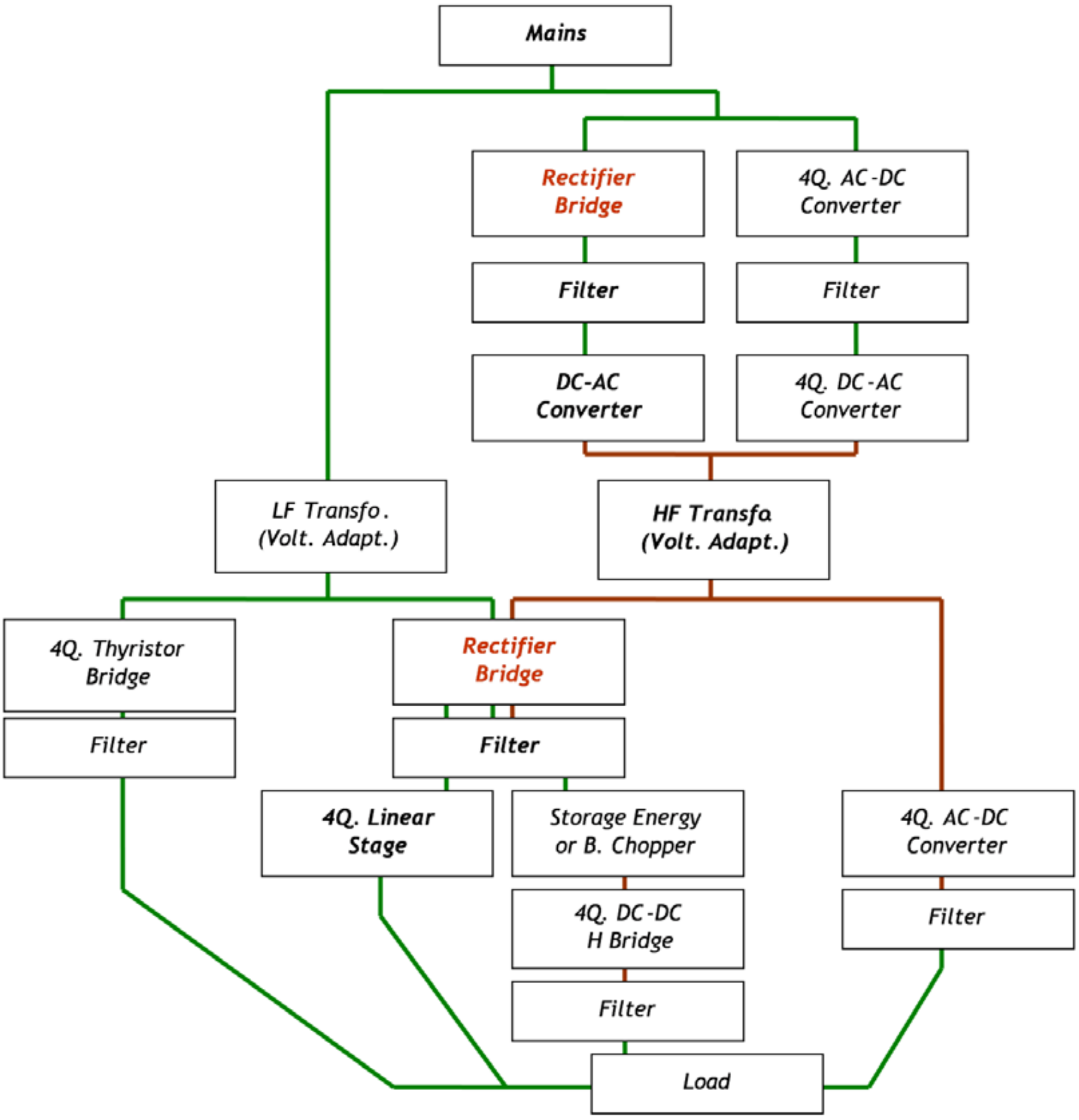

**Fig. 12:** Overview of possible combinations

The second part of this paper describes the design and realization of a power converter using a linear stage (indicated in bold in Fig. 12).

## 5 LHC120A-10V power converter design

This part describes the practical realization of a [±120 A; ±10 V] four-quadrant power converter, designed at CERN in 2003–2004 for LHC use.

### 5.1 Description

The converter was designed for being integrated into a high performance environment: its main use is to provide parts-per-million precision current to a superconducting magnet. A low level of EMC perturbations created by the power converter, on the input and output sides, and a high level of conformance toward the reference voltage to be followed (no distortion, high bandwidth) are strong requirements for the high precision electronics located in the power rack. Its integration in the existing tunnel is a natural constraint, which means that only switching base topologies are adequate: high efficiency minimizes the losses that must be evacuated from underground installations; and a small size and low weight are required for installation and repair interventions around the 27 km of the LHC tunnel. A long lifetime is also required from a system that will operate for more than 10–15 years.

### 5.2 Schematic overview

The converter is based on a high switching frequency phase-shifted standard topology, in series with a four-quadrant linear stage, and can be represented as shown in Fig. 13.

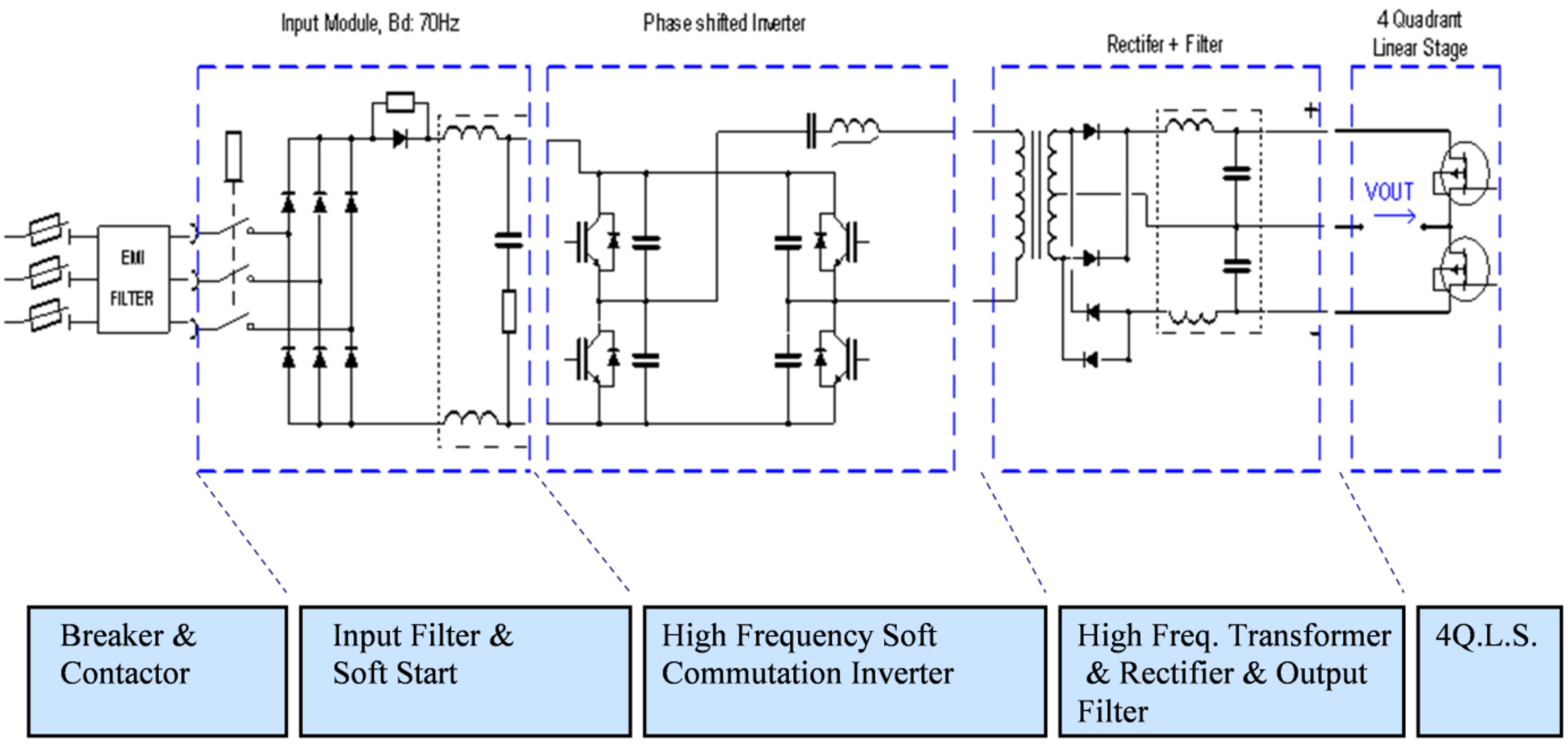

**Fig. 13:** LHC120A-10V schematic overview

Figure 13 shows that the power converter is in five parts.

- Protection and power control stage: a breaker is used for protection and safety reasons when an AC contactor is used to power the power converter or isolate it from the mains.
- AC–DC stage: a conventional rectifier bridge and input filter (70 Hz) with soft start capability provides DC voltage to the next stage.
- High frequency DC–AC inverter: a 70k Hz phase-shifted zero voltage switching (ZVS) inverter with insulated-gate bipolar transistors (IGBT) switches DC voltage into a high frequency voltage given to the power transformer in charge of adapting and isolating the output side.

– Isolation and rectifier stage: a high frequency power dual output transformer and low voltage Schottky power diodes output high frequency dual DC voltage to the four-quadrant linear stage.

– Four-quadrant linear stage: based on power MOSFET transistors (mounted in parallel on each side to boost the receiving energy capability), capable of absorbing and dissipating the entire load energy. This stage's performance is not dependent on a quadrant being in operation, and it has additional functions: minimum load of previous stage, active filter, and high rejection of the mains natural and expected 300 Hz ripple present at the AC–DC stage.

### 5.3 Four-quadrant linear stage

The main element of this power converter is the four-quadrant linear stage. In particular, the choice of power MOSFET for this stage leads to a specific design, to handle the inherent limitations of this kind of component (threshold, non-linear component).

#### 5.3.1 *Principles*

Figure 14 shows a different working example, as a first approach to the system regulation laws.

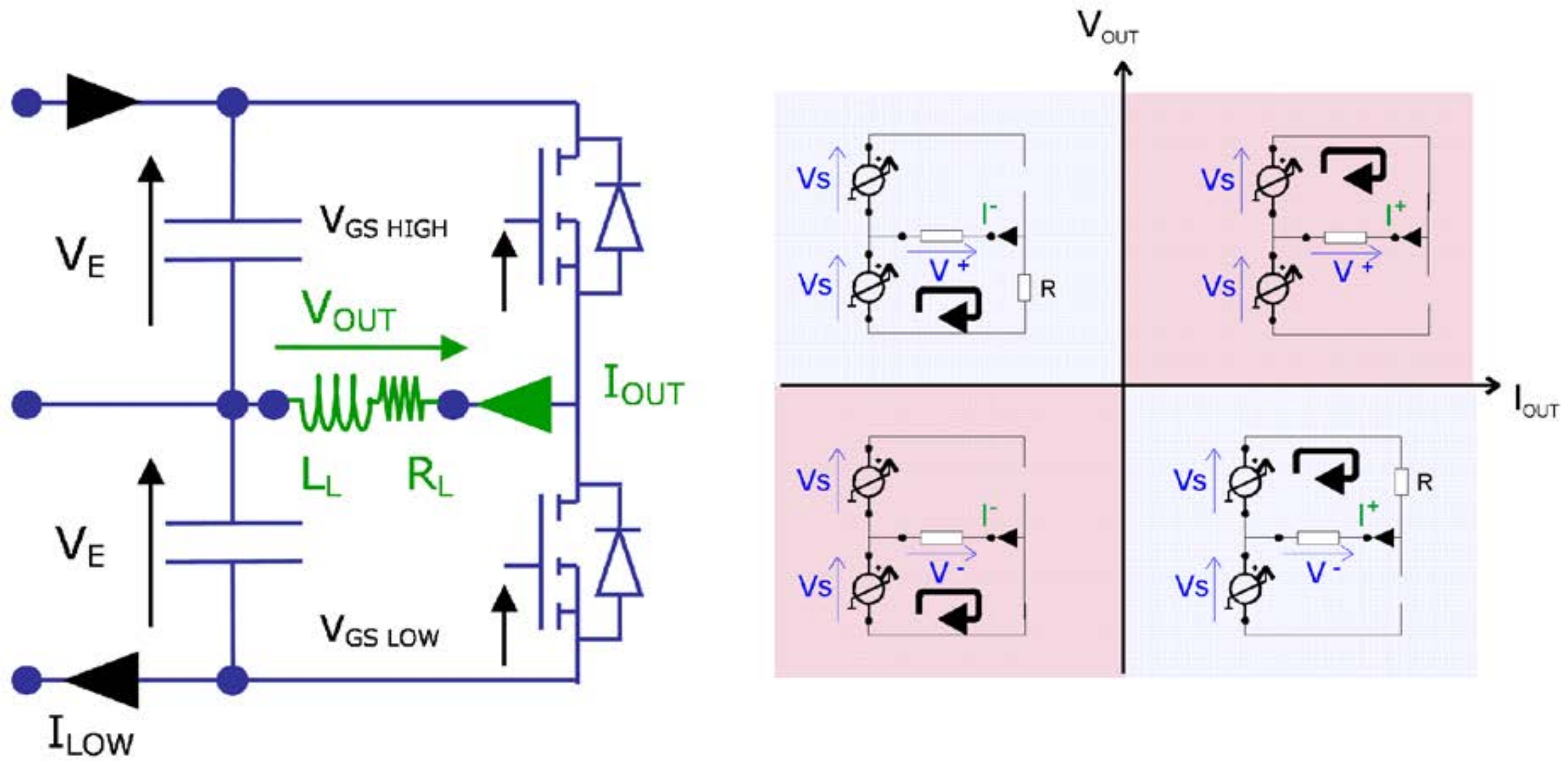

(a) (b)

**Fig. 14:** Four-quadrant linear stage schematic (a) electrical schematics; (b) operational areas

The control principles can be simplified to two equations (Eqs. (3) and (4)), each of which are valid for two quadrants:

$$V_{\mathrm{OUT}} = +\left(V_{\mathrm{E}} - R_{\mathrm{HIGH}} \cdot I_{\mathrm{HIGH}}\right) \text{ Quadrants 1 and 2} \quad (3)$$

$$V_{\mathrm{OUT}} = -\left(V_{\mathrm{E}} - R_{\mathrm{LOW}} \cdot I_{\mathrm{LOW}}\right) \text{ Quadrants 3 and 4} \quad (4)$$

where $R_{\mathrm{HIGH}}$ and $R_{\mathrm{LOW}}$ are the equivalent power MOSFET resistances.

## 5.3.2 *Power MOSFET characteristics*

[redacted]*Power MOSFET, a natural current source*

A power MOSFET is a natural current source, since the gate voltage determines the current flowing into it, independently from the voltage seen by it. This can be easily extracted from the manufacturer's data in Fig. 15, where a specific gate voltage leads to a constant current versus drain-to-source voltage $V_{DS}$. (Reference: FB180SA10, International Rectifier in this example below).

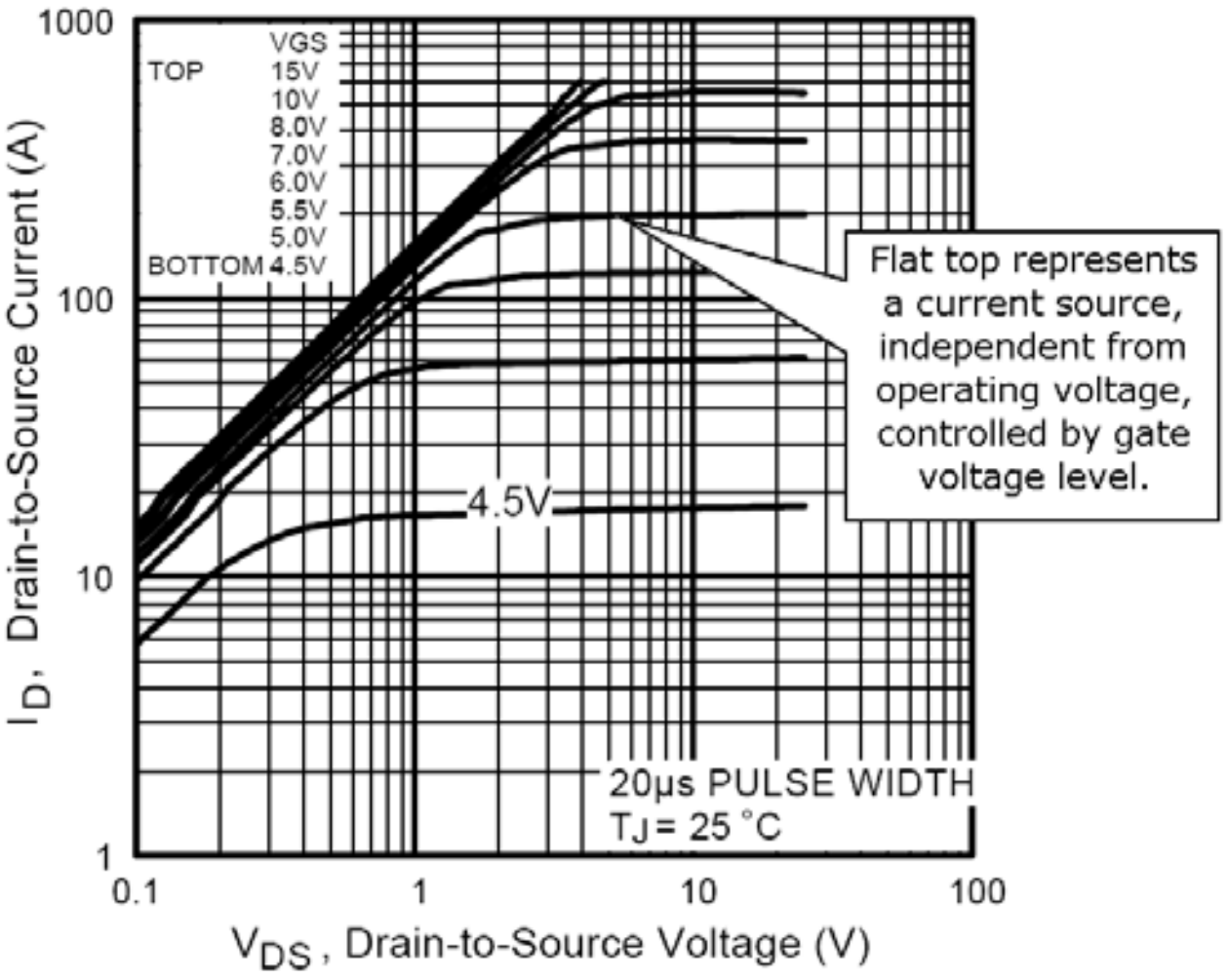

**Fig. 15:** Typical power MOSFET characteristics

The nature of the load involved, the superconducting magnet (current source per excellence), requires an output decoupling filter to be able to connect both the current source and the current load. Moreover, values of the load time constant involved in this practical realization were in a range that was so large that this option was not retained. A complete linear stage was nevertheless produced and successfully tested, using the intrinsic current source of a power MOSFET, being used in the flat part of its curve.

[redacted]*Power MOSFET transistor, a 'controlled resistance'*

A power MOSFET transistor used as a 'controlled resistance' is a highly non-linear system. The following curves summarize the main phenomena that should be taken in account: influence of temperature, $V_{DS}$ voltage, Miller capacitance, etc. Of course, the internal structure of a power MOSFET should be taken in account, especially with modern power MOSFETs, which are more able to switch than work in a linear mode.

[redacted]Main parameters to be taken in account

Some key parameters are highlighted, always focusing on the gain $G$:

$$G = \frac{R_{\mathrm{DSON}}}{V_{\mathrm{GS}}}. \qquad (5)$$

[redacted]Gain variation over working range

A classical $R_{DSON}$ versus $V_{GS}$ curve shows that several orders of magnitude of static gain should be considered, where a linear model around an operating point is used. If a linear control system is chosen, the working range of the power MOSFET should be selected to minimize gain variations. Figure 16 shows a typical curve for a standard power MOSFET.

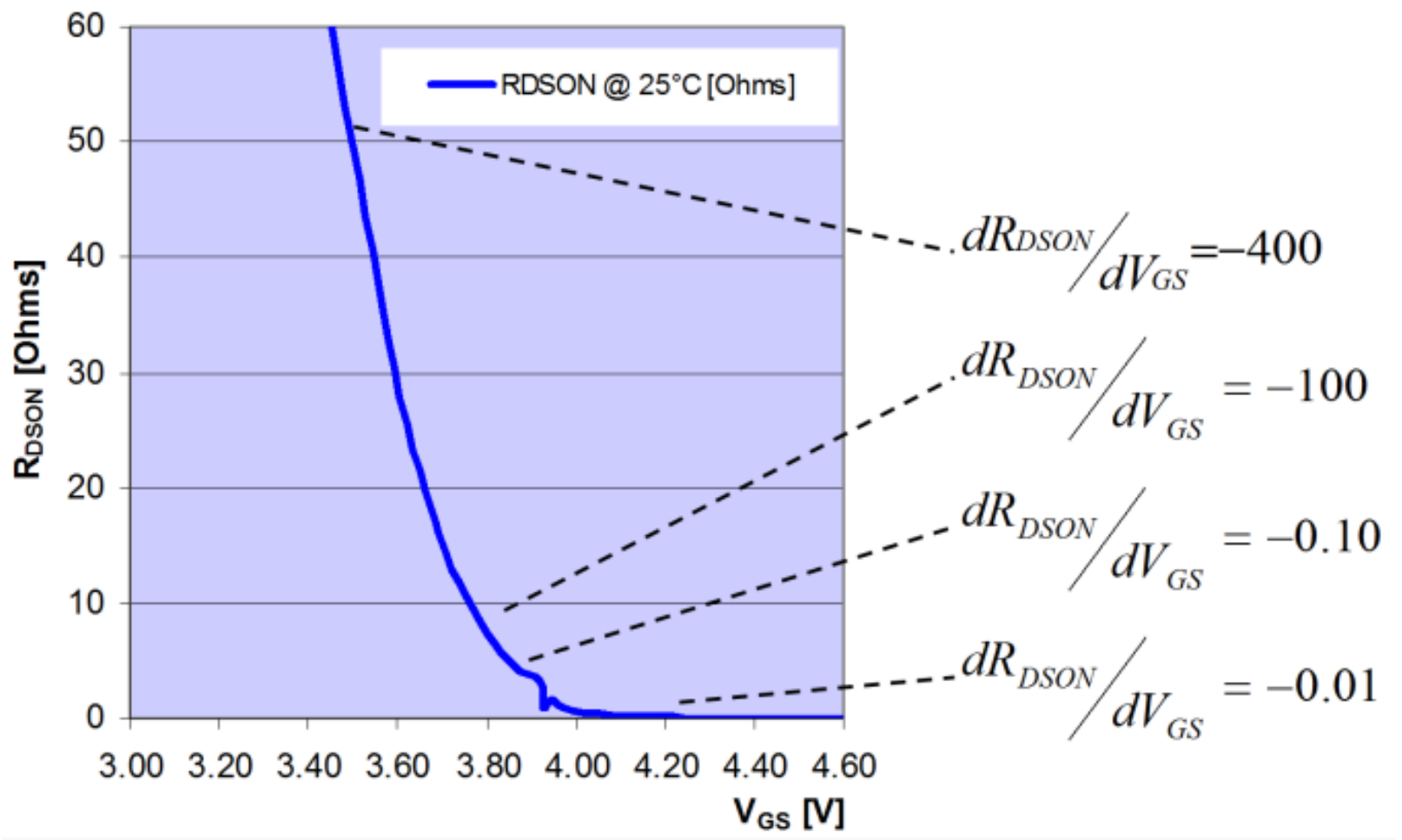

**Fig. 16:** MOSFET $R_{\mathrm{DSON}}$ overview

[redacted]Threshold variations over a batch

The conduction threshold is a key parameter, since a transistor linear stage cannot expect a step voltage control signal that is driven by the linear control main loop to suddenly switch a power MOSFET into conductive mode without providing the correct $V_{\mathrm{GS}}$ threshold voltage. The knowledge of this threshold variation is mandatory when trying to cope with it. Manufacturers give a large range for this parameter since it can vary from 2–4 V. This means that a fixed threshold to approach the linear conductive mode is not a valid approach and cannot be considered for a safe and robust design.

[redacted]Threshold variations and radiation

Variation of the threshold gate voltage versus radiation dose is a key factor for some converters used in particle accelerators. Figure 17 shows the results for a power MOSFET [1000 V; 15 A] threshold variation versus radiation dose; this power MOSFET is used in a test configuration to control a constant low current (50 mA).

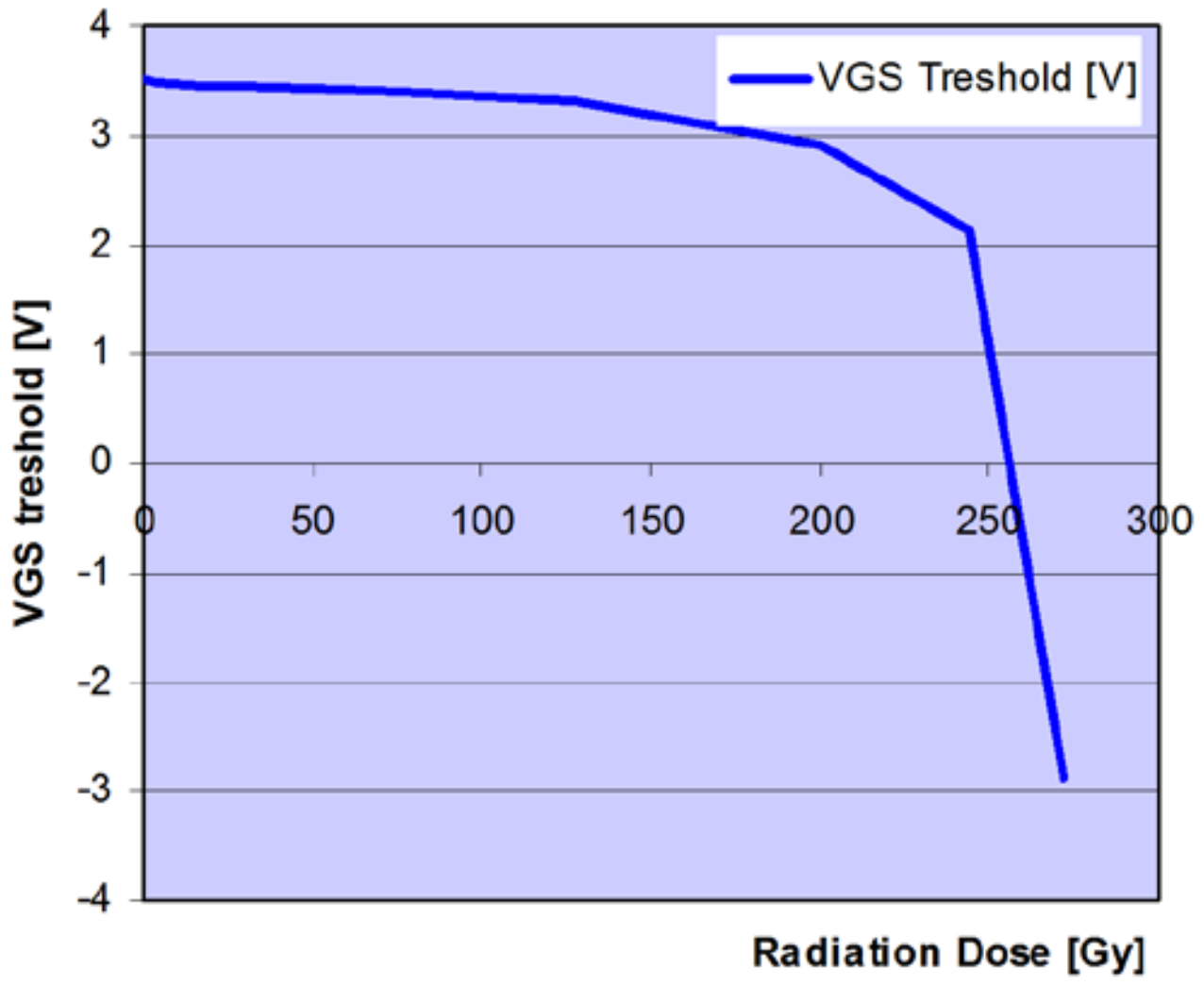

**Fig. 17:** Power MOSFET $V_{\mathrm{GS}}$ threshold variation versus radiation dose

Considering single-event sensitivity, the fact that a power MOSFET can stay in a continuous conductive state (even with a very low current being flowing through the transistor), with the voltage across it being lower than the maximum capability, makes this topology almost non-sensitive to single-event failures.

### [redacted]Temperature influence

The influence of temperature is well-known for power MOSFET transistors, with an $R_{DSON}$ that increases with temperature. When controlling a power MOSFET transistor in linear mode, increasing the temperature introduces a negative threshold ‘offset’ as described in Fig. 18.

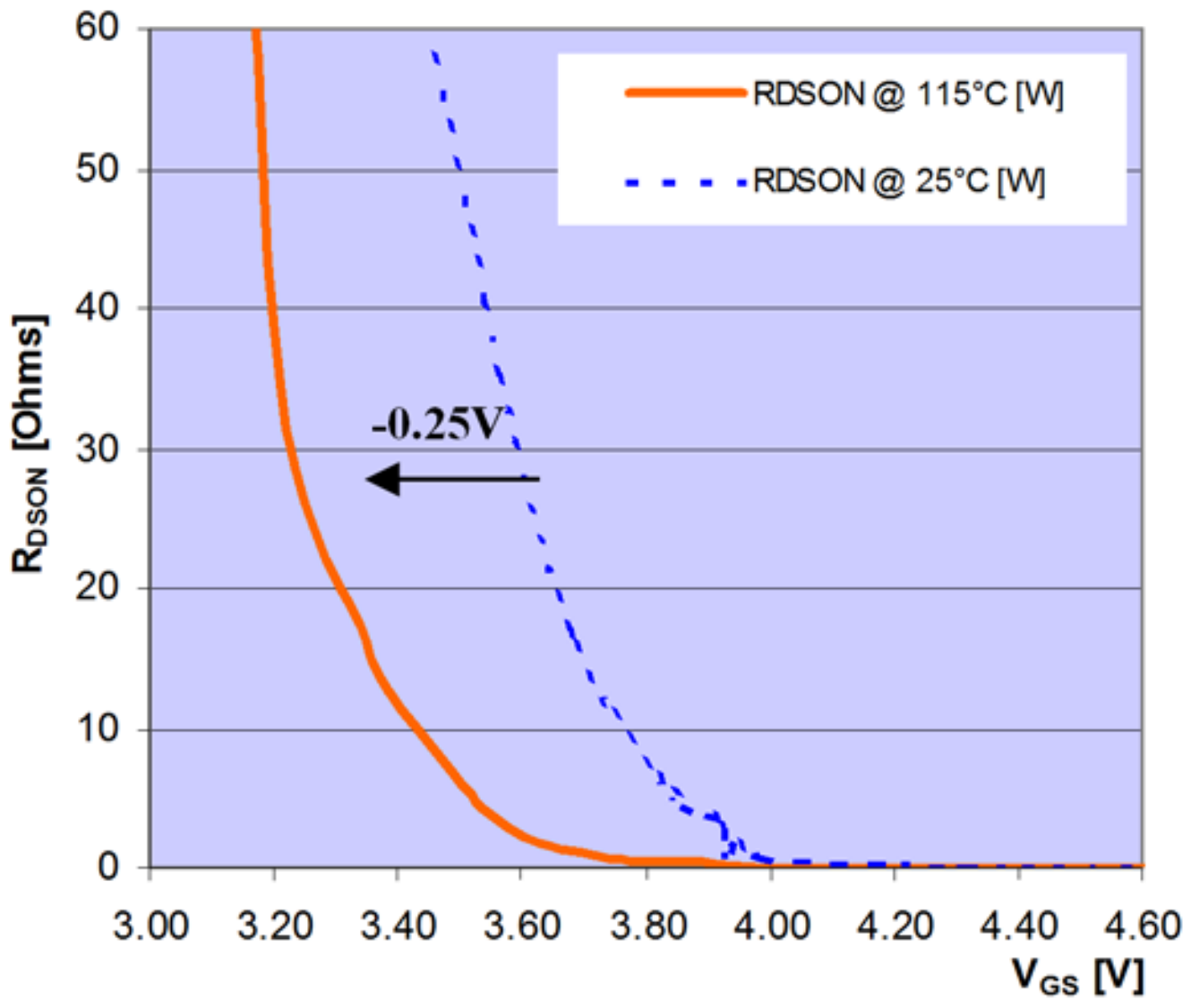

**Fig. 18:** MOSFET $R_{DSON}$ versus temperature

This feature is important since, again, it prevents the use of a fixed threshold, *even* if trimmed per power MOSFET at 25 °C, for each power MOSFET threshold level.

### [redacted]$V_{DS}$ influence

Gain is also influenced by the voltage across MOSFET $V_{DS}$. See the curves in Fig. 19. This curve has to be used to determine the minimum value of $V_{DS}$ for operation. Indeed, in a classical linear stage, we will see that if maximum voltage is determined by the load operation area, the minimum value across a power MOSFET can be selected based on efficiency and control criteria. It can be observed that choosing a low operating voltage across the power MOSFET will lead to a square angle characteristic with a high rate of change that is difficult to control.

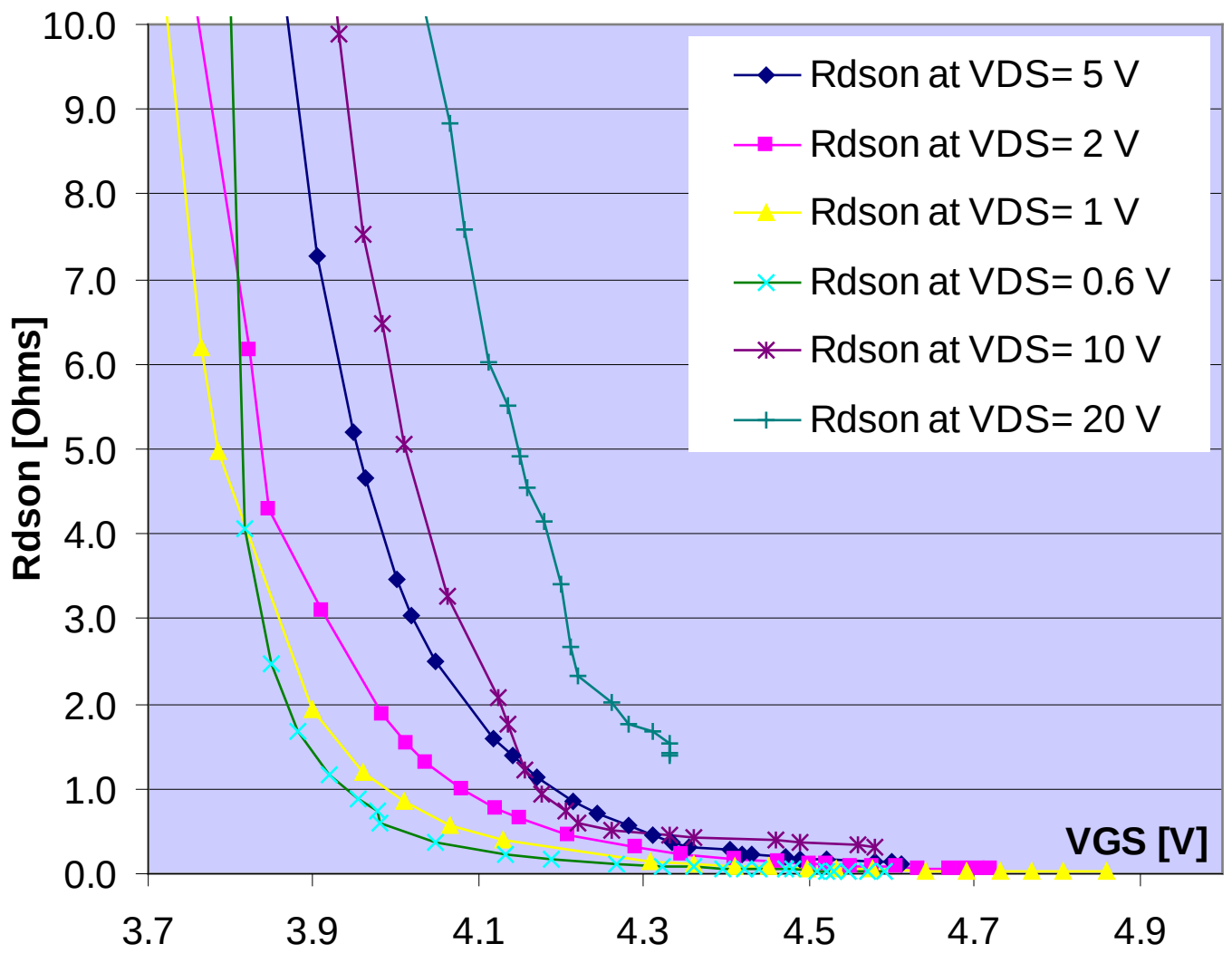

**Fig. 19:** Power MOSFET $R_{DSON}$ versus $V_{DS}$

[redacted]Miller capacitance effect

A signal applied to the gate voltage is loaded by a well-known capacitance that changes with the power MOSFET conduction status. This change could affect the bandwidth of the control signal if used with a high resistance in series with a power MOSFET gate. Total gate capacitance is a combination of the proper gate capacitance added to the output capacitance depending on the power MOSFET status: if a power MOSFET is conducting, and close to its minimum $R_{\mathrm{DSON}}$, gate capacitance seen from the gate can be multiplied by a factor up to 2–3. In linear mode use, this capacitance can easily be derived from the manufacturer's data:

$$Q_{\mathrm{g}} = C_{\mathrm{iss}} \cdot \left(V_{\mathrm{GS\ Final}} - 0\right) + C_{\mathrm{iss\ Miller\ add}} \cdot \left(V_{\mathrm{GS\ Final}} - V_{\mathrm{GS\ Threshold}}\right) \tag{6}$$

where $Q_{\mathrm{g}}$ is the total gate charge, $C_{\mathrm{iss}}$ is the grid capacitance, $V_{\mathrm{GS\ Threshold}}$ is the power MOSFET conduction threshold, and $C_{\mathrm{issMilleradd}}$ is the Miller capacitance.

From this equation can be found the capacitance range:

$$V_{\mathrm{GS}} \leq V_{\mathrm{GS\ Threshold}} \rightarrow C_{\mathrm{iss}} \ , \tag{7}$$

$$V_{\mathrm{GS}} \geq V_{\mathrm{GS\ Threshold}} \rightarrow C_{\mathrm{iss}} + C_{\mathrm{iss\ Miller\ add}} \ . \tag{8}$$

[redacted]*Power MOSFET model conclusions*

Controlling a power MOSFET like a 'variable resistance' presents two major problems when using a linear system: a conduction threshold on the gate applied voltage, and a high difference of 'static' gain depending on the resistance value to be obtained.

### 5.3.3 *Power MOSFET push–pull control*

A linear loop is used to control the power MOSFET push–pull, based on an opposite signal from the control loop sent to the gate voltage of the power MOSFET transistor branches. In that configuration, the control loop has to provide a threshold voltage added to the small signal control signal once the power MOSFET reaches the linear zone. It can be noticed that when switching from the upper to the lower leg, the control signal has to provide a step of two times the gate voltage threshold.

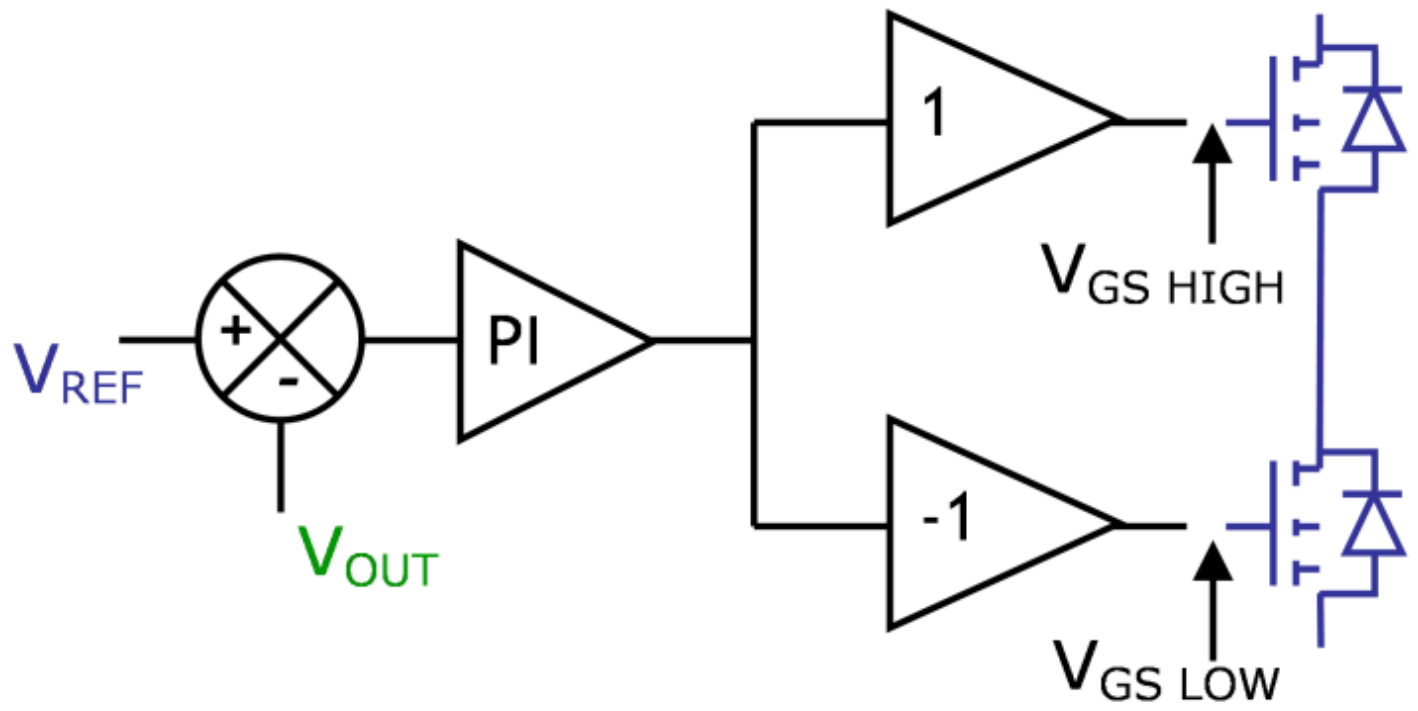

**Fig. 20:** Push–pull control principles

Even if this is highly non-linear due to the power MOSFET's $R_{\mathrm{DSON}}$ behaviour, this principle works quite well when taking in account the high level of $R_{\mathrm{DSON}}$ when used with a low current, while a high level of current requires a low value for $R_{\mathrm{DSON}}$. Indeed, a small signal analysis would show that it is possible to choose a working voltage across the power MOSFET (design choice) so that each compensate themselves, giving an acceptable gain variation.

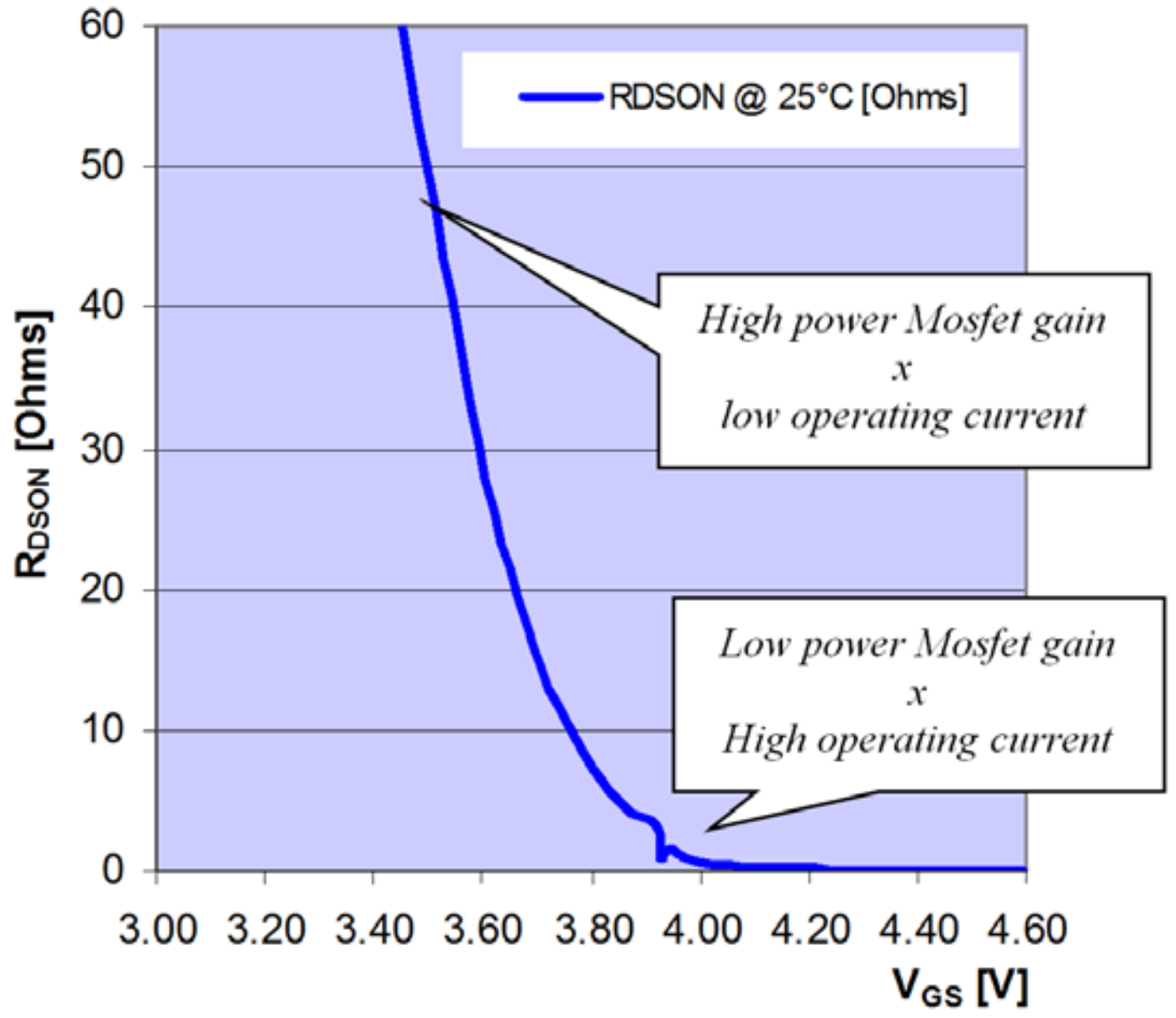

**Fig. 21:** Small signal analysis detail

### 5.3.4 *Inherent limitations of a push–pull stage*

Even if it is very simple, a basic push–pull system has an uncontrollable zone at null current, since voltage cannot be 'obtained' from the current in the load if it is null. This problem is enhanced with a superconducting load, where full voltage can be apply with null current. In the same order, the [0 V; 0 A] point is an unstable point. When applied to a power MOSFET stage, two severe limitations appear.

- Gate voltage threshold, which makes it difficult to use a simple linear system (a control signal given by the loop has to switch between $+V_{\text{GS threshold HIGH}}$ and $-V_{\text{GS threshold LOW}}$ as fast as possible to avoid the output becoming uncontrolled (blank area, where no power MOSFET conducts). Figure 22 shows the step voltage required from the voltage control loop; up to 8 V in this example (two times 4 V from each power MOSFET threshold level).
- Use of power MOSFET equivalent resistance area can be very wide, almost saturated to an almost open state, giving a four decade gain variation, which again is very difficult to control in a simple way, even with the self-compensating effect of the current being related to $R_{\text{DSON}}$ values.

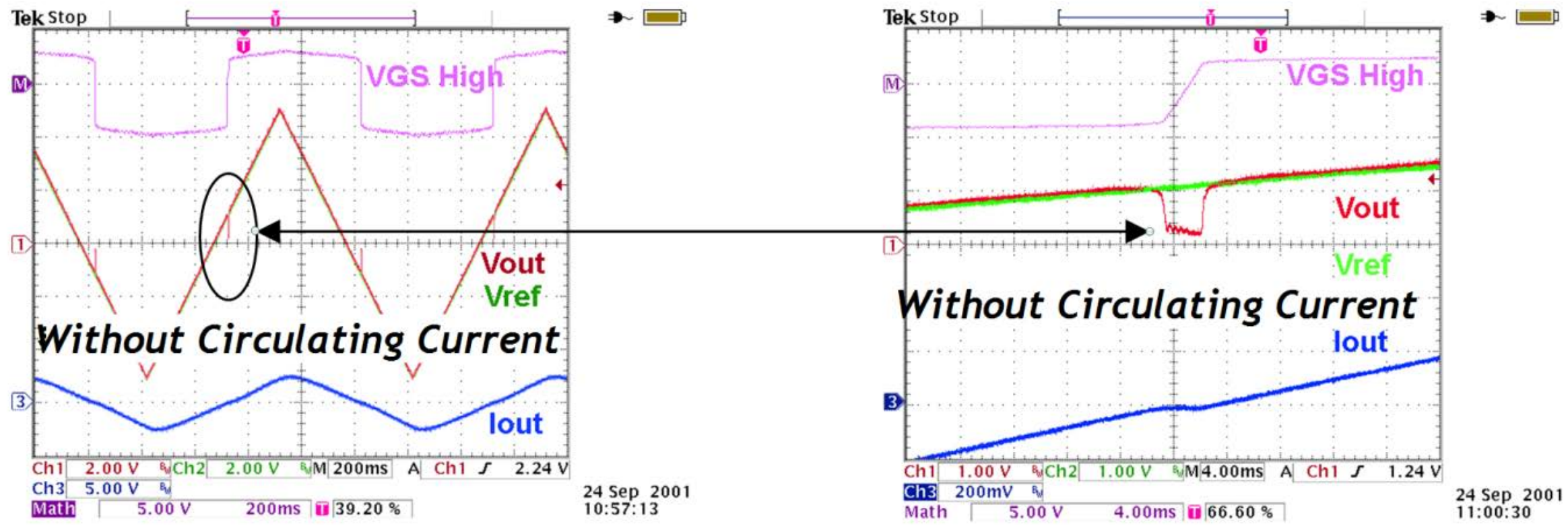

(a) (b)

**Fig. 22:** MOSFET-based push–pull stage limitations: (a) cycle overview; (b) zoom on 0 A crossing transition

### 5.3.5 *Circulating current*

A circulating current is a controlled current, internal to the converter – not seen by the load – that maintains a known state for both the DC–DC power converter and the four-quadrant linear stage. This feature was requested to obtain high level performance. It provides the following functions.

- Provides a minimum load for the DC–DC power converter and avoids a completely non-loaded output side of the dual output DC–DC power converter. The power DC–DC switched-mode converter is therefore easily controllable, without a problem of management of continuous and discontinuous mode, and there is a clamping voltage on the output capacitors on each side, whatever the output conditions.
- Pre-conditions the linear stage power MOSFET, close to the threshold gate voltage. The linear stage loop will be able to manage high-side and low-side transitions since none of the power MOSFETs requires a voltage step to change mode (conductive or not).
- Limits working zone of the power MOSFET used in linear mode. It is possible to redefine artificially the range of operation of power MOSFET of each leg in linear mode playing with the value of the circulating current combined with the operating voltage.
- Less important but very convenient is the capability of having a current circulating inside the power converter when trimming a power MOSFET, if a sharing procedure has to be used; this is when using power MOSFETs in parallel per side. A power converter can be trimmed without any external load, with an adjustable level of current being set to optimize this possible procedure.

Figure 23 shows a simple representation of the two loops involved in that mode. The addition of this loop – circulating current – provides real advantages, and gives possibility to – dynamically – trim the system very deeply (in this case the value of the circulating current is controlled depending on an adjustable limit on the output current). It is nevertheless an additional loop, which should not disturb the two loops already in place: power MOSFET push–pull and power DC–DC control. Great care of the bandwidth of this loop is mandatory, and high signal dynamic performance of the overall power converter are partly deduced from this loop's characteristics.

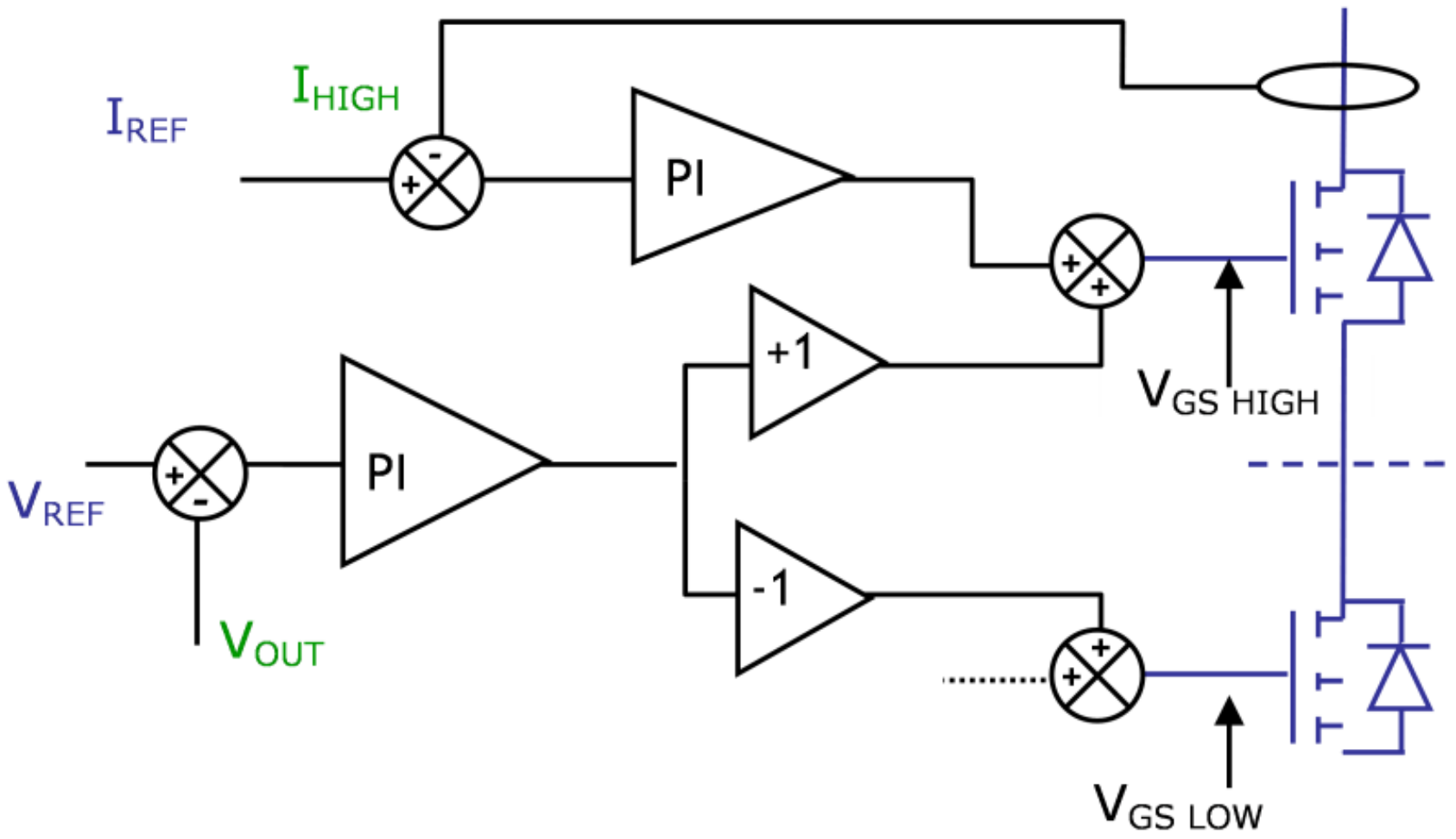

**Fig. 23:** Push–pull control principle with circulating current loop

#### [redacted] *Influence of circulating current on operational range*

Circulating current can be used to limit drastically the highly non-linear zone to a more practicable one, limiting the high gain zone. This is particularly interesting since it doesn't cost too much in efficiency, reducing considerably the less controllable linear zone of the power MOSFET (high resistance, high gain), which would lead to instability. Avoidance of this problem is especially critical with a superconducting load, where the power converter maximum output voltage can be applied in a receiving quadrant while current, even if slightly moving, is very low.

In the case where a linear stage provides full voltage at a current close to zero, a power MOSFET will be used with a high resistance value, inducing a very high gain, and becoming a critical zone for linear loop stability. The circulating current strategy ensures minimum current flows in each power MOSFET, which are then capable of producing the desired output voltage with medium values of $R_{DSON}$, while not just relying on output current. Figure 24 shows a drastic reduction of the range of $R_{DSON}$, using only 1 A of circulating current.

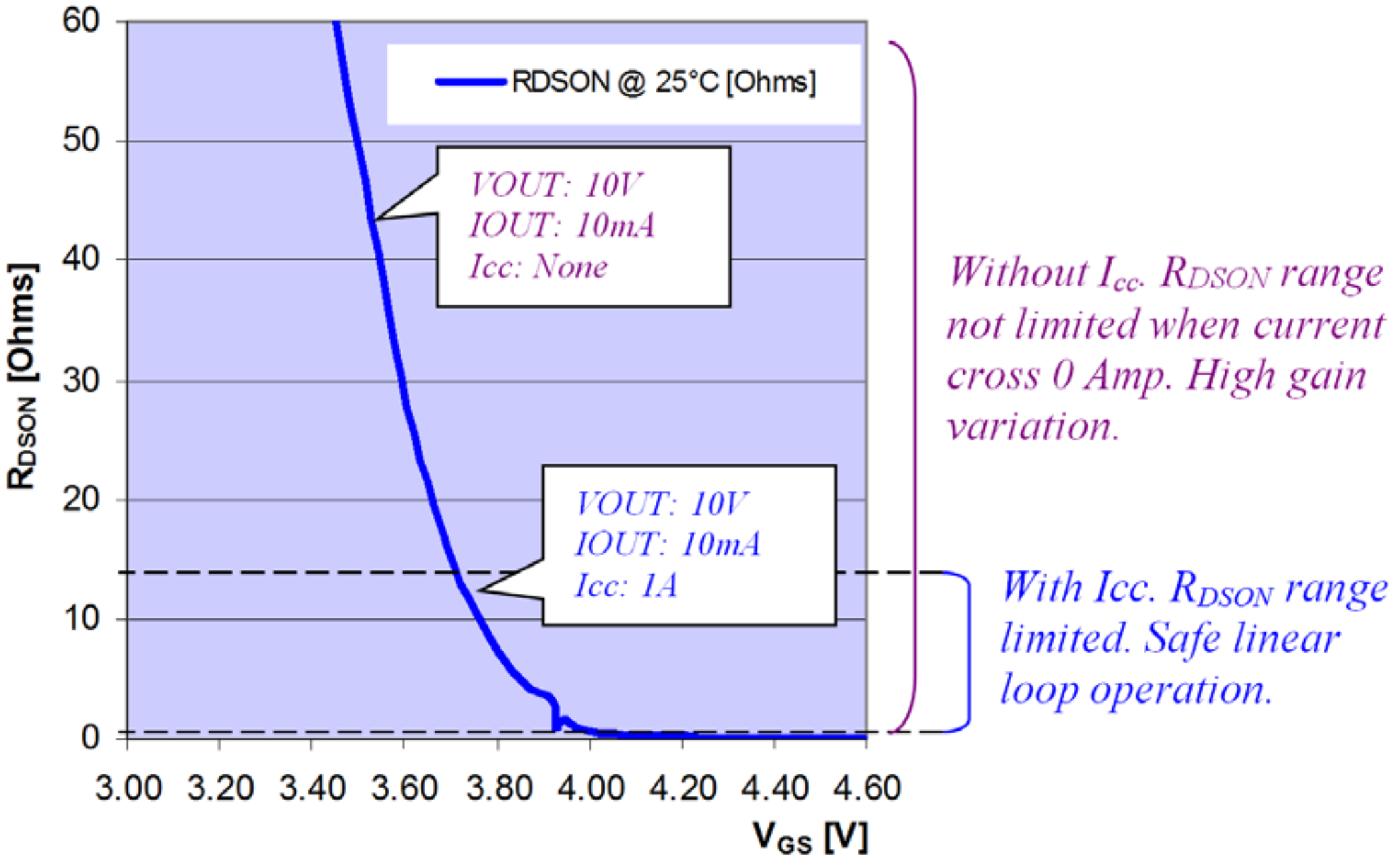

**Fig. 24:** Small signal analysis detail

### [redacted] *Circulating current implementation*

Implementing this feature requires the addition of a current transducer per side. (Note that these two sensors can be used to deduct the output current of the load, which is often required for load protection.) In such a case, the following can be noted.

- Knowing the output current will make it possible to adjust the circulating current level to improve overall efficiency.
- Even if these sensors are of low precision (percentage level), they should be of a sufficiently high bandwidth that they do not interfere too much with the inner loop.
- It is possible to use shunts in series with each power MOSFET to provide a self-damping behaviour if the gate voltage is applied to a power MOSFET gate and measurement shunt.
- It is mandatory to use a higher reading range than the power converter produces, since each sensor will alternatively, and depending on the current polarity, see the output current added to the circulating current at a maximum output current if non-null (depending on the circulating current strategy). It is nevertheless possible to use a circulating current only around the 0 A area, making it possible to use a low-current sensor. This cost-effective solution will limit the potential benefit of the addition of a circulating current loop.

### [redacted] *Circulating current results*

Figure 25 shows the result of a circulating current on gate voltage, being constantly close to the threshold voltage of each power MOSFET.

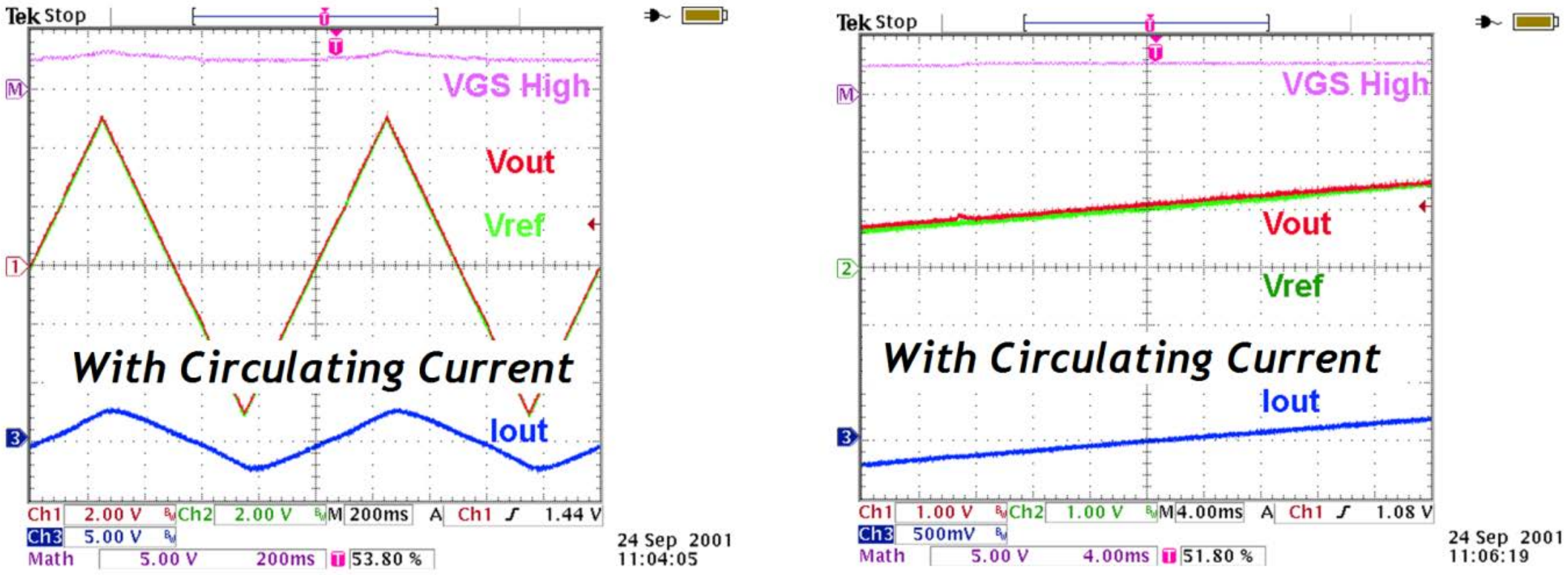

(a) (b)

**Fig. 25:** Circulating current improvement effect: (a) cycle overview; (b) zoom on 0 A crossing transition

### [redacted] *Circulating current possible operation*

Circulation current leads to losses, especially when the output conditions are far from 0 A. The worst example is full output voltage, where circulating current causes the largest amount of energy to be lost in a power transistor. Considering that this circulating current is only useful close to 0 A output current, different strategies can exist, with $I_{LIMIT}$ close to 0 A output current, to avoid extra losses.

A first possible strategy, with circulating current only present close to transitions, is given below:

– $I_{OUT} < I_{LIMIT}$ $I_{CC} \rightarrow$ $ON_{high\ value}$ Operating close to 0 A output current is possible;

– $I_{OUT} > I_{LIMIT}$ $I_{CC} \rightarrow$ OFF Circulating current removed above a threshold.

Another possible strategy, with circulating current always present, is given below:

– $I_{OUT} < I_{LIMIT}$ $I_{CC} \rightarrow$ $ON_{high\ value}$ Operating close to 0 A output current is possible;

– $I_{OUT} > I_{LIMIT}$ $I_{CC} \rightarrow$ $ON_{low\ value}$ Circulating current is always present, even if the value is reduced.

The $I_{CC}$ low value is determined so that high frequency dual output voltage DC–DC is sufficiently loaded on the non-leading side to avoid overvoltage. It can be noticed that a very low level of $I_{CC}$ is sufficient to polarize both power MOSFETs, and therefore avoid a too long delay before entering the safe 0 A crossing zone.

Figure 26 shows the positive effect of a non-null circulating current on the capability of dealing with a high current change rate. If the two operating cases seem almost equal for operating close to 0 A output current, the result is nevertheless different when taking into account the speed of the circulating current loop. Indeed, this additional loop will be lower than the main controlling loop and stabilization time can become critical if $dI/dt$ is too fast. If this can occur, the second solution is a lot better. This is particularly interesting when the converter has to work on a pure resistive load, where $dI/dt$ is not limited by the load.

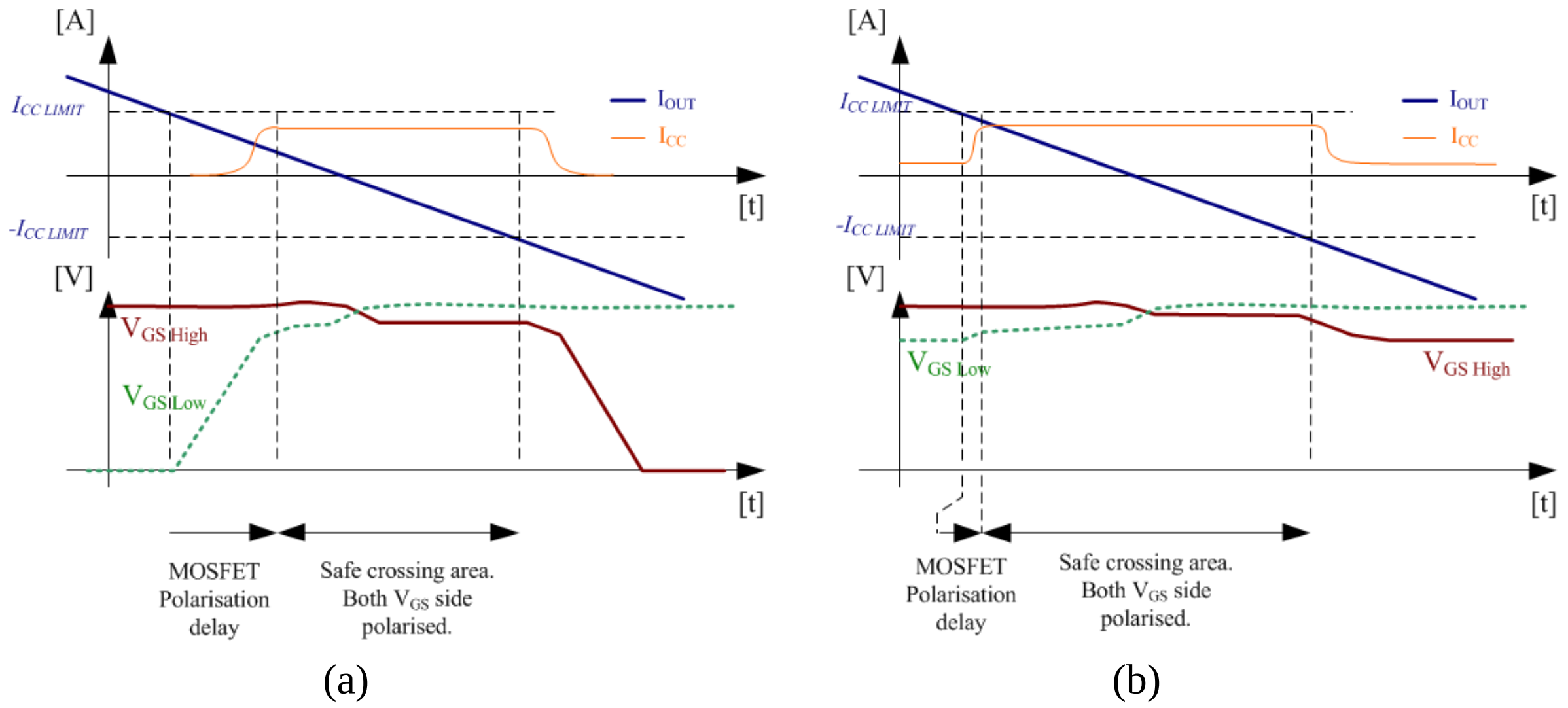

**Fig. 26:** Effect of circulating current on power MOSFET gate polarization: (a) circulating current set to null value case when not required; (b) circulating current being always non-null, with value changing according to the output current value.

### [redacted] *Influence of the number of power MOSFETs on a linear loop*

The minimum number of power MOSFETs to be used is determined by power or, better, by the energy to be absorbed by the linear stage. It is obviously possible to use a large number of power MOSFETs so that the $V_{BIAS}$, $V_{DS}$ voltages across MOSFETs in generator mode can be decreased, while keeping a reasonable controllable area for the power MOSFETs when used at a high current in generator mode.

Reducing $V_{BIAS}$ makes it possible to increase efficiency, at the price of additional power MOSFETs. Nevertheless, this improvement is balanced by the fact that MOSFETs will work in a higher gain area, due to the $R_{DSON}$ characteristics curve, when absorbing energy from the load, leading to a potential instability. If this design feature can easily be taken in account, increasing the power of an existing converter does not just require an increase in the number of output power MOSFETs; control loops have to be reworked to ensure proper operation.

### [redacted] *Medium and high frequency rejection results*

As noted above, a power MOSFET is a natural current source, and it will offer a natural rejection of both medium and high frequency, from DC–DC dual output to four-quadrant linear stage (4QLS) output. If rejection is part of the control loops, the medium and high frequency rejection feature is clearly of benefit to the active power MOSFET characteristics.

Figure 27 shows voltage measurements from a real converter; $V_{OUT}$ is the output voltage measurement, when $V_{E+}$ is the voltage provided by the DC–DC power part. Measurements were performed at maximum power, to stress the input power filter regarding the 300 Hz expected content.

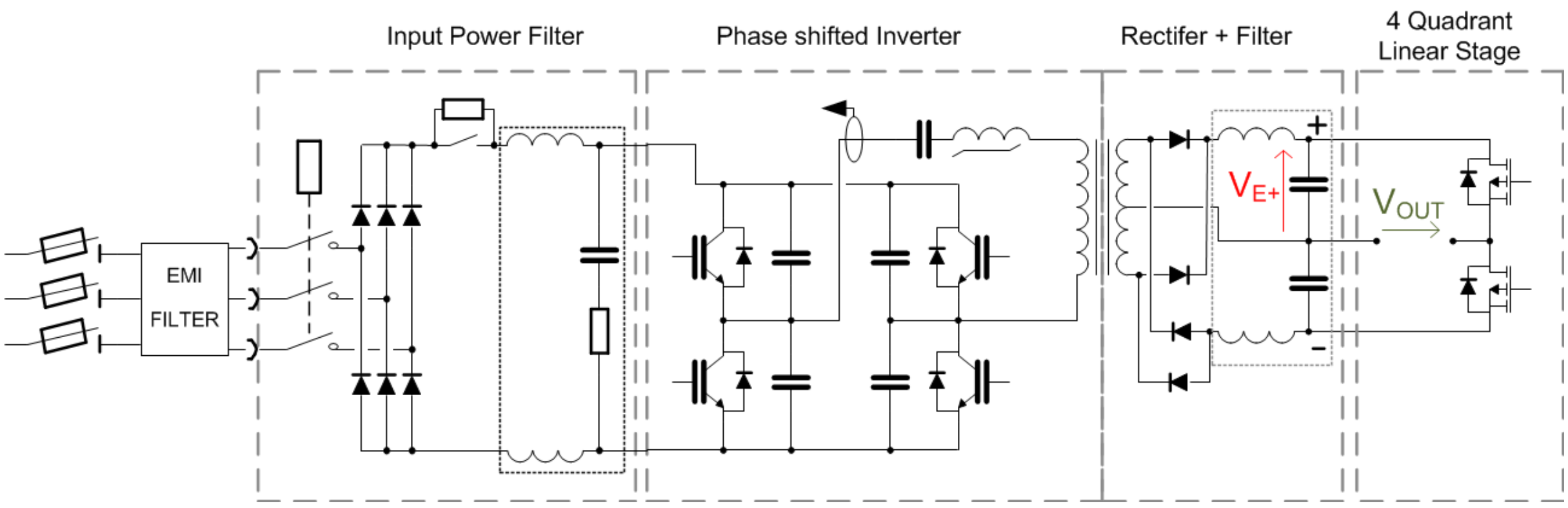

**Fig. 27:** Schematic with voltage measurements of interest

Figure 28 indicates that medium frequency rejection is very high; the performance comes from the voltage loop natural rejection, related to its bandwidth. The result is a dramatic reduction of the 300 Hz by a factor of 15. Decreasing $V_{BIAS}$ will decrease the rejection of medium frequency since the gain of the power MOSFETs will be lower, being close to their saturation area (flat and low gain). A trade-off between the efficiency and the level of noise at 300 Hz is to be evaluated. Figure 29 shows the direct reduction effect of the power MOSFET current source behaviour in the high frequency range.

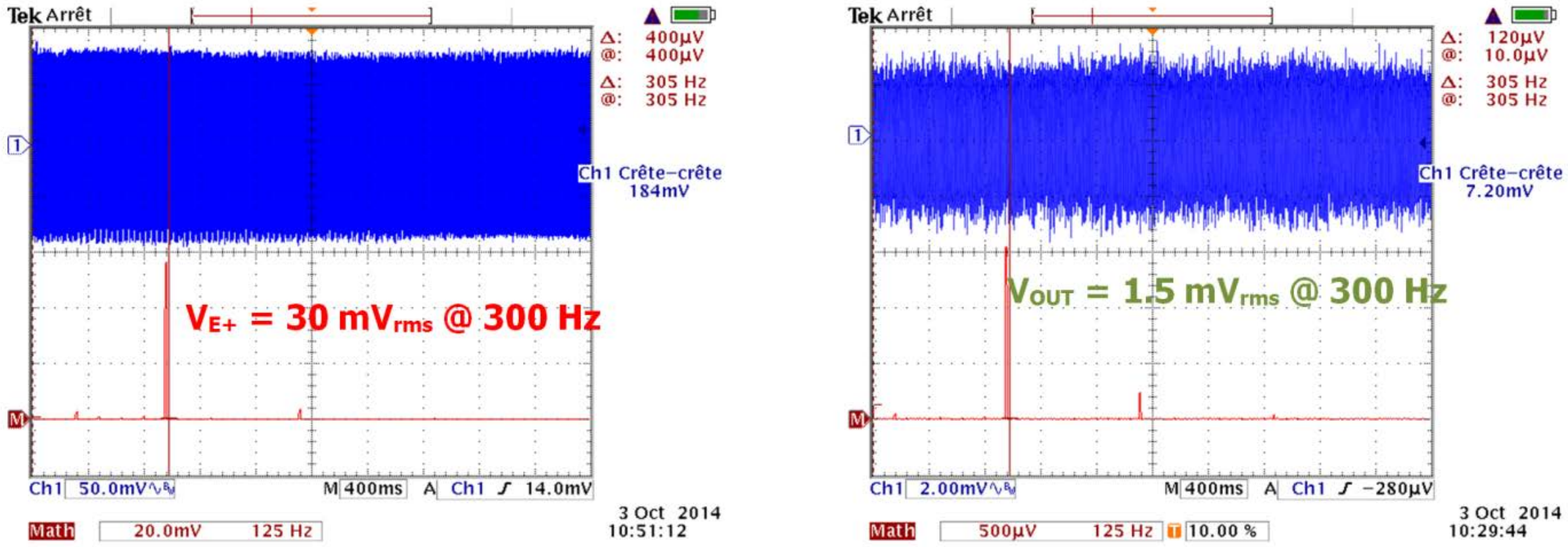

(a) (b)

**Fig. 28:** 4QLS medium frequency reduction measurements on a real converter: (a) 300 Hz level on DC intermediate Bus; (b) 300 Hz level on the output voltage.

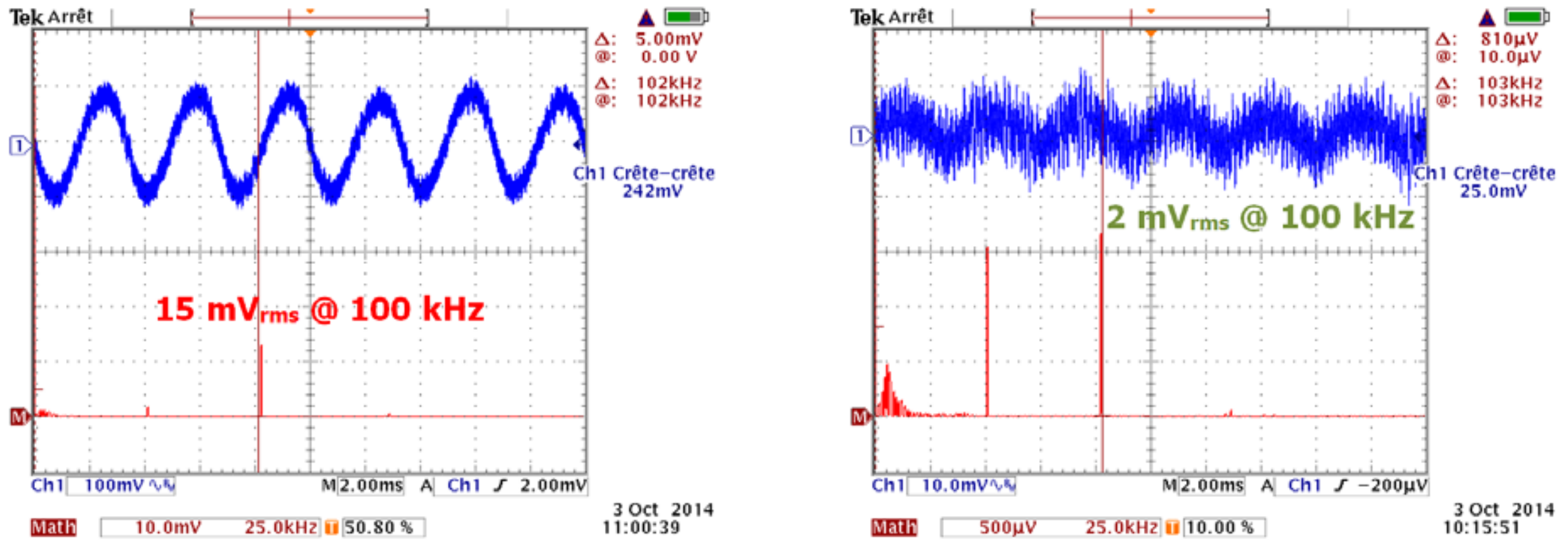

(a) (b)

**Fig. 29:** 4QLS high frequency reduction measurements on a real converter: (a) high frequency level on DC intermediate Bus; (b) high frequency level on the output voltage.

## 5.4 Converter control loops

Since a four-quadrant linear stage is fed by the power DC–DC (inverter plus dual output filter), a control strategy for the power DC–DC, four-quadrant linear stage, and circulation current loops has to be decided upon. Since cascaded loops are involved, a factor of ten is considered between power DC–DC and four-quadrant linear stage loop speeds. The 4QLS is the fastest loop, since the generating and absorbing modes both using the 4QLS, but not the generating one. Indeed, 4QLS is fed by the load in quadrants two and four, and power DC–DC can be transparent in that mode.

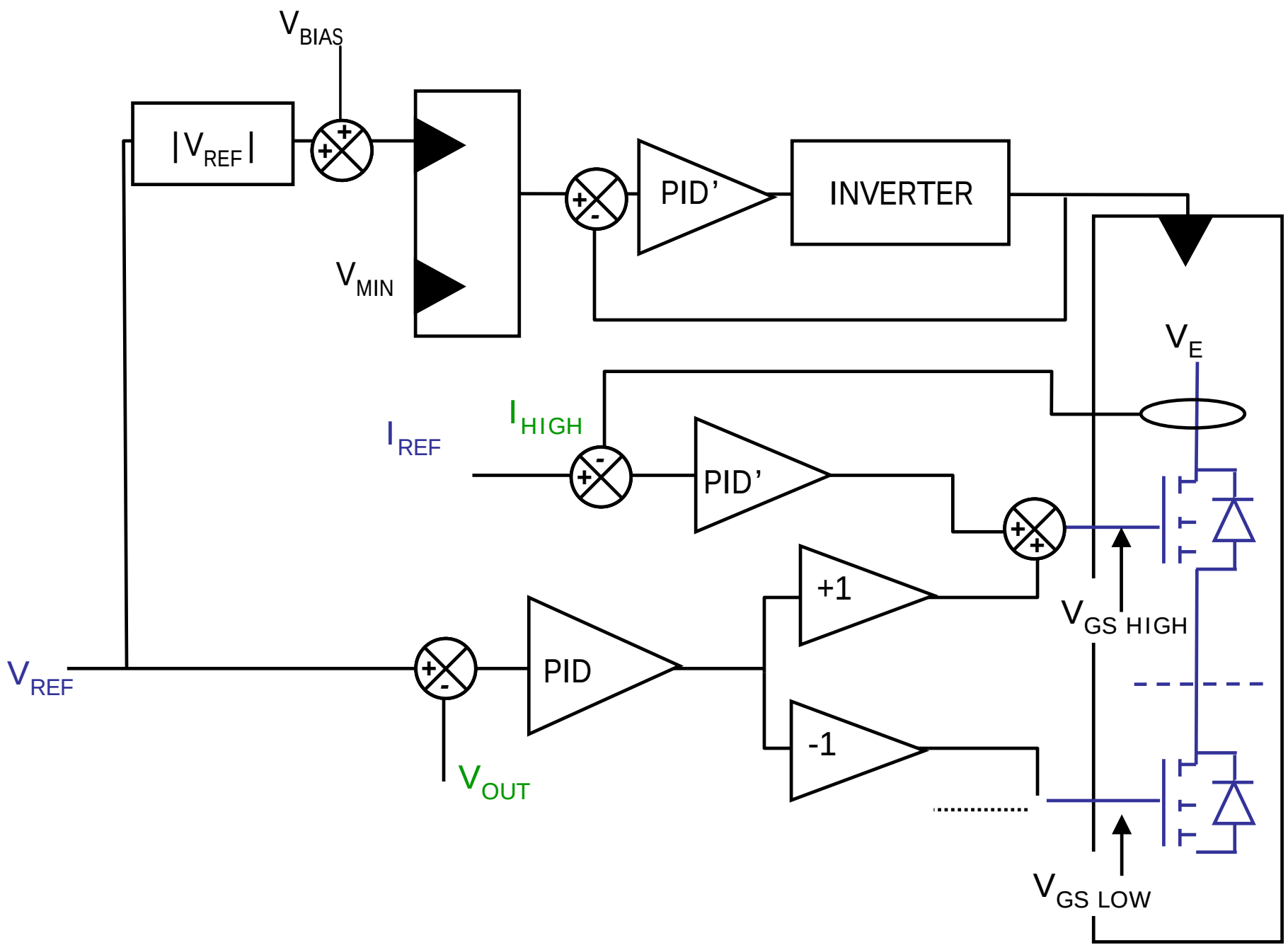

**Fig. 30:** Overall control strategy

Circulation current is always the slower loop, since polarization of the power MOSFET should be completely transparent from the other loops, and being assimilated to perturbations to be rejected by the other loops. The current rate can give a minimum speed for this loop to achieve non-distorted crossings.

### 5.4.1 *Power DC–DC loop*

This loop is at least ten times slower when compared to the four-quadrant linear stage. A voltage reference coming from the converter user was chosen to be the same bandwidth as the power DC–DC, to avoid the four-quadrant linear stage becoming faster than what the power DC–DC could provide, leading to conflicts. In this case two settings are used trim the power DC–DC reference: $V_{\mathrm{BIAS}}$ and $V_{\mathrm{MIN}}$. The first is used so that the four-quadrant linear stage can work in a given operating range (not too saturated) as described in previous chapters. The second is used to ensure that power DC–DC is always at minimum loading by circulating current and a minimum given output voltage reference, leading to a minimum power. It also makes the loops more robust close to [0 A, 0 V], limiting the operating zone where all three loops are active at the same time. See Fig. 31.

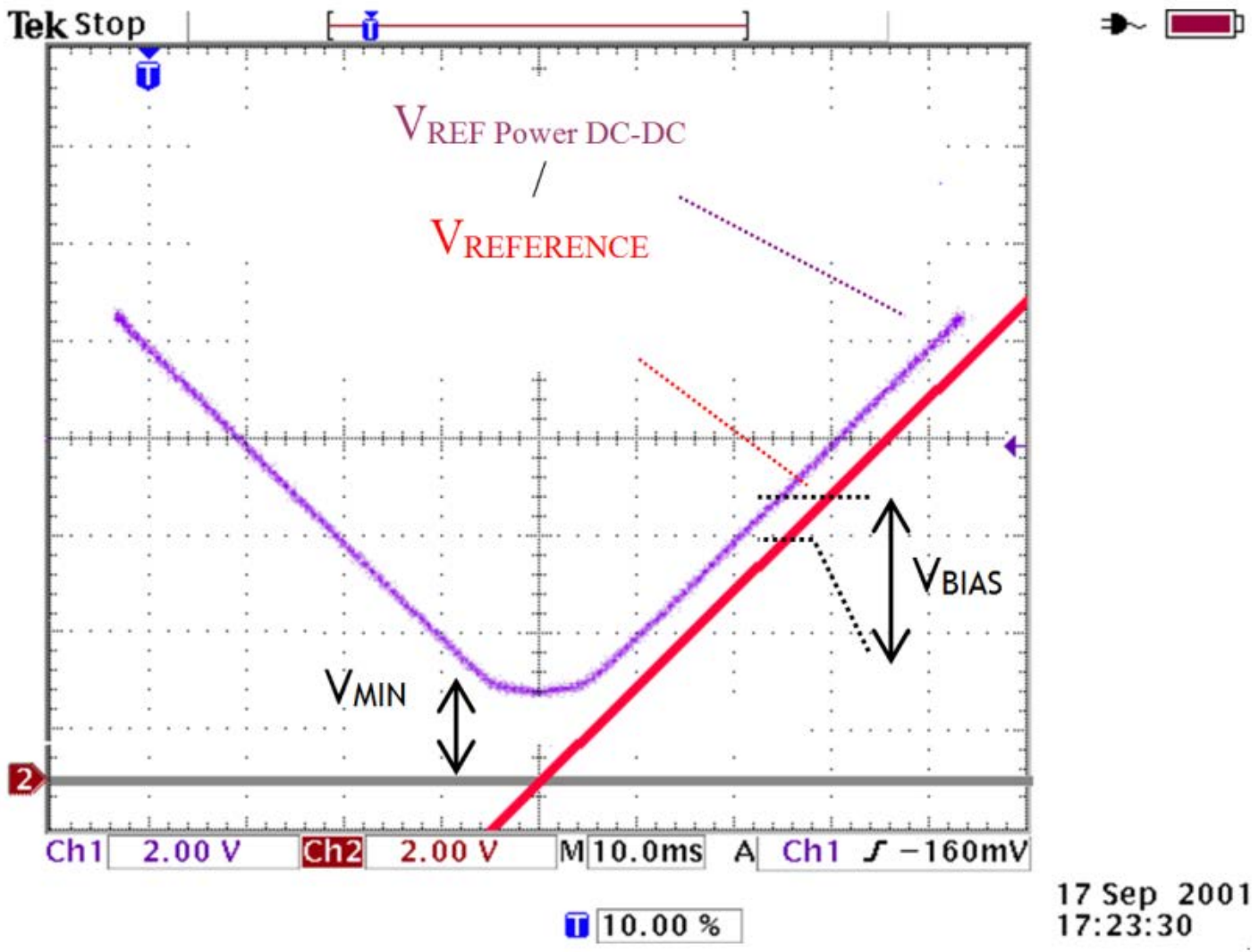

**Fig. 31:** Typical power DC–DC reference versus converter reference

## 6 Practical results

The converter was manufactured by the company Efacec, Moreira da Maia, Portugal, in 2004–2005, which was also in charge of the rack design and the industrialization of the proposed solutions, following the CERN design. The converter module is housed in a 5U 19″ 45 kg power module, as shown in Fig. 32. It includes control signal, interlocks, and signalling capabilities on the front of the module.

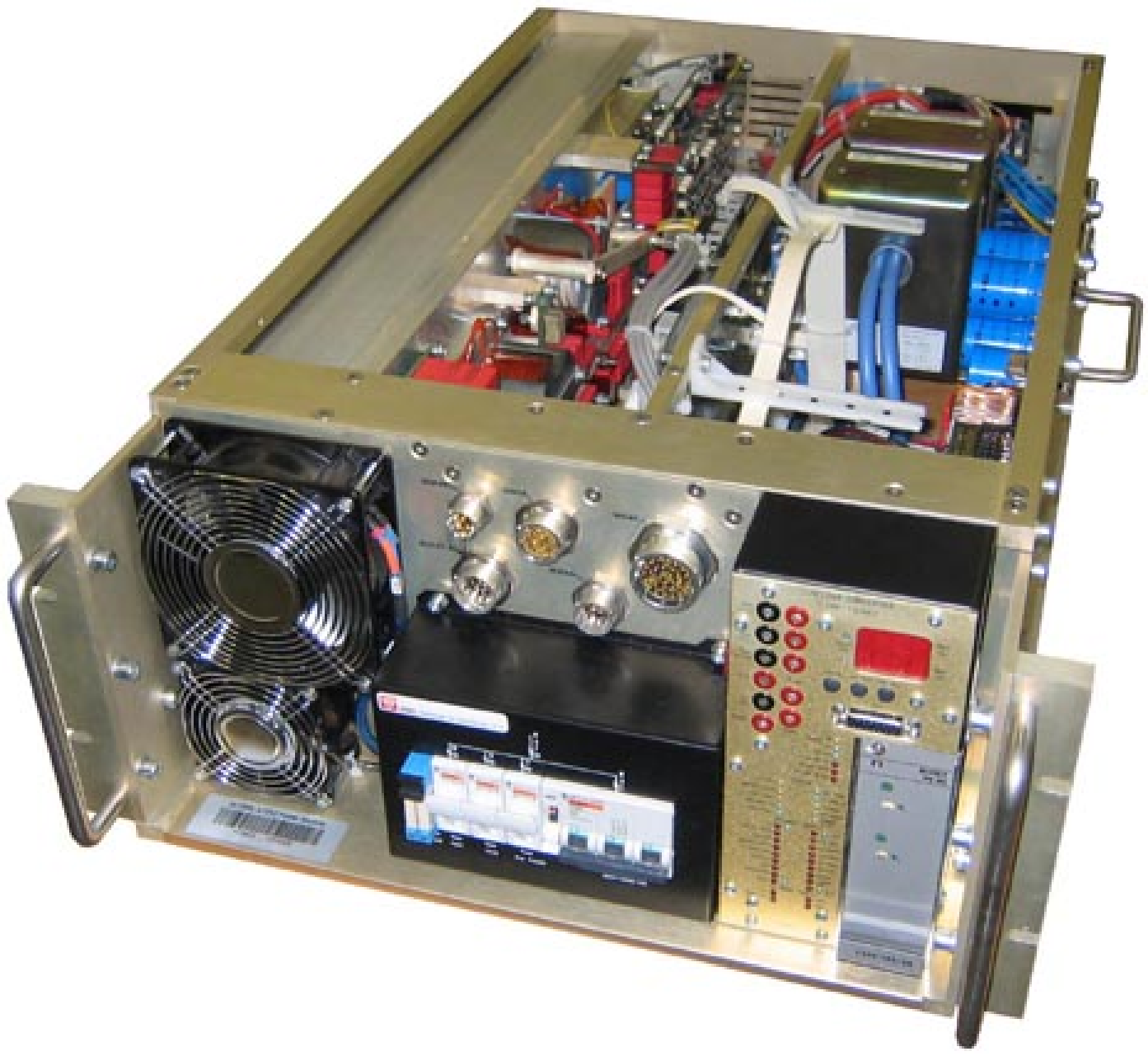

**Fig. 32:** LHC 120 A-10 V power module

Power converter characteristics during production confirmed the initial choice of the topology. The system is robust, with a bandwidth of 1 kHz, without any distortion of the voltage while crossing the 0 A point, and the EMC level was confirmed to be very low. The different trimming possibilities ($V_{BIAS}$, $V_{MIN}$, circulation current levels (low and high level), and its current limit when changing level) and the cascade type design made adjustments easy to achieve, since the system is not too interleaved. Efficiency was measured to be relatively low (72% at full power) due to linear stage losses.

Figure 33 shows a 0 A crossing plot, without any distortion, for different loads.

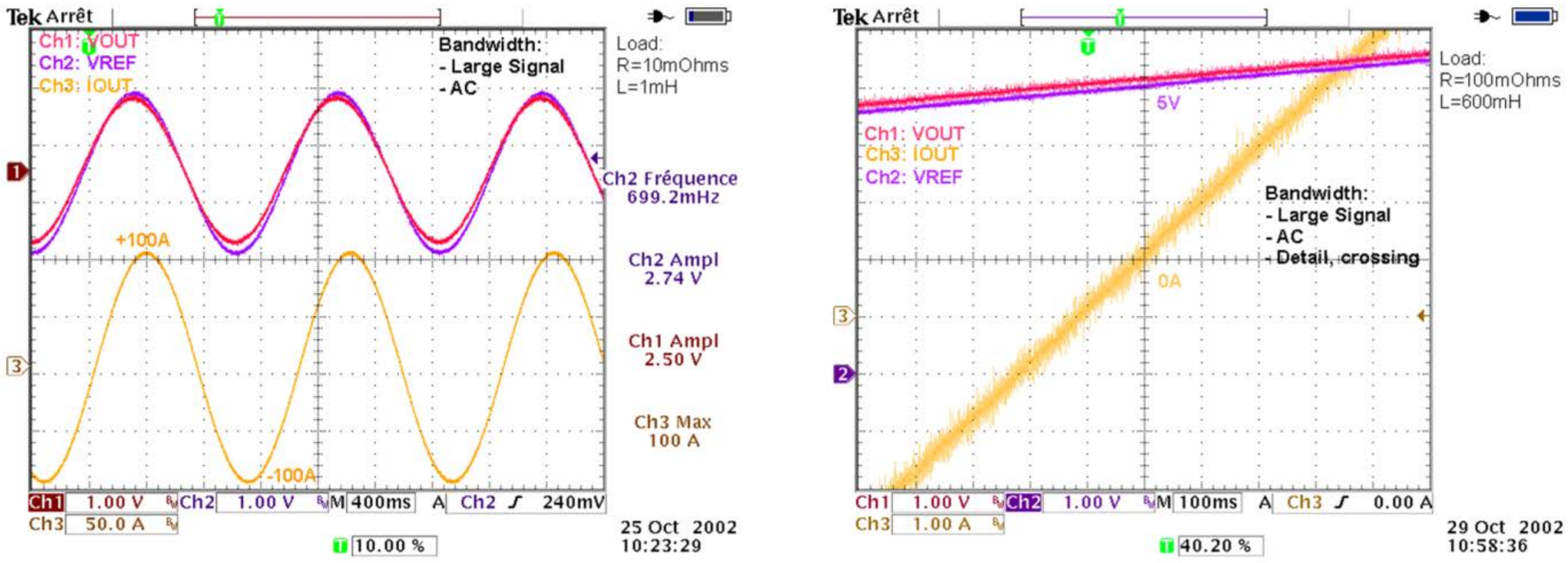

(a) (b)

**Fig. 33:** Crossing 0 A measurement for different magnet loads: (a) 0 A crossing performance on low time constant magnet; (b) detail of a 0 A crossing on large time constant magnet.

Moreover, the bandwidth of the system is constant whichever quadrant is used, ensuring safe operation on a superconducting magnet in regard to the very high precision level (some parts per million) of the current loop, which will not be affected by the load operating conditions. Figure 34 shows a typical plot of generator and receptor bandwidth. A voltage step is required, with a small signal added, so that the converter operates in generator mode, then receptor mode. The current doesn't change at this timescale, since the superconducting magnet keeps it unchanged (large time constant)

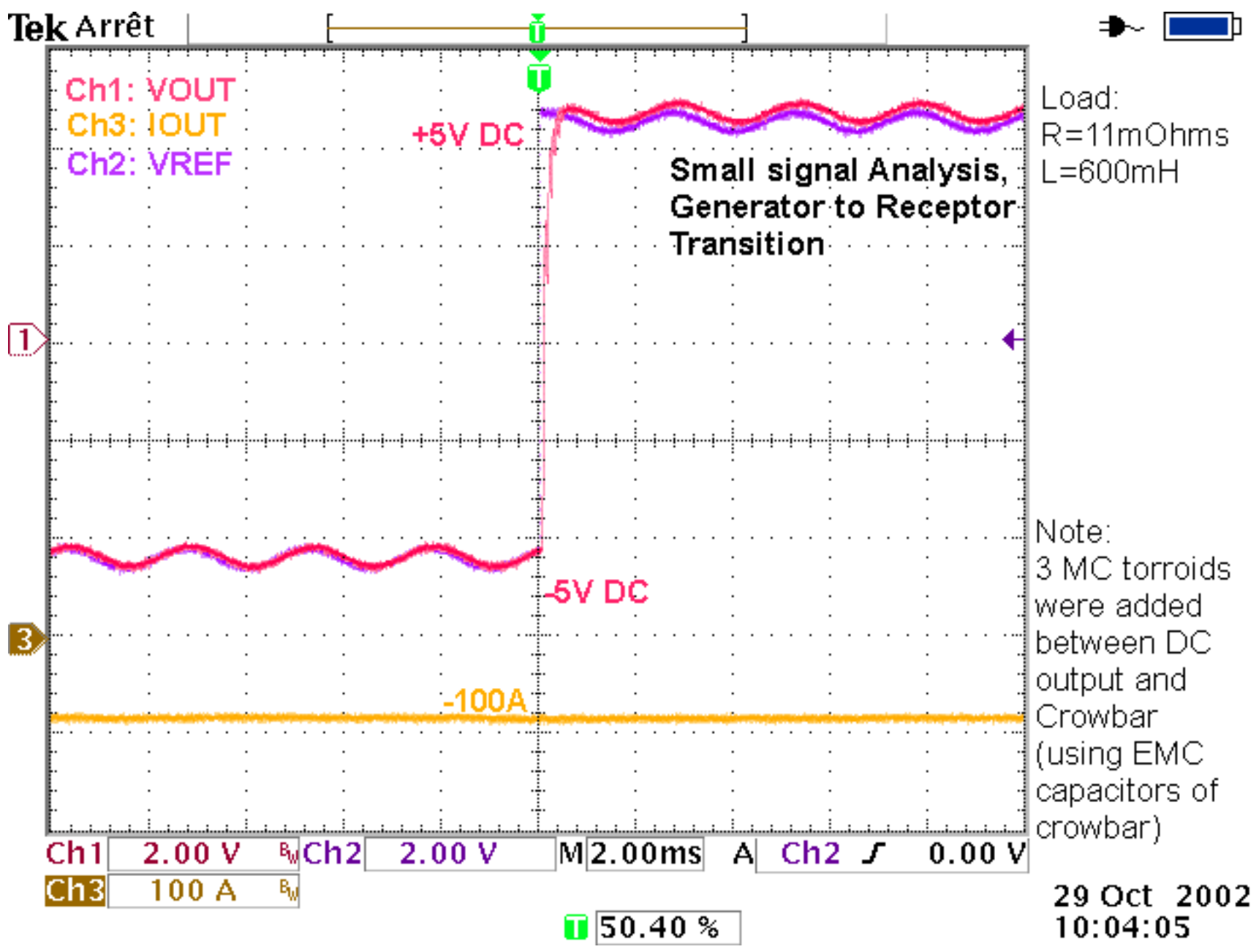

**Fig. 34:** Small signal analysis in receptor quadrant

EMC levels were measured on the AC side (input) and the DC side (user side, superconducting magnet in series over hundreds of metres). If the AC level is low, according to the IEC regulation standards, the DC side was found to be extremely low, especially in the high frequency range. This result is mainly due to the soft commutation inverter, added by the one switching stage alone topology. This choice was made possible to easily deal with parasitic components (limiting common and differential mode capacitors of transformers and secondary side rectifiers) using output filter as an additional EMC component. The linear stage is obviously a quiet element that doesn't affect converter EMC, in opposition with a second switching stage. See Fig. 35 for a typical plot at maximum current and voltage.

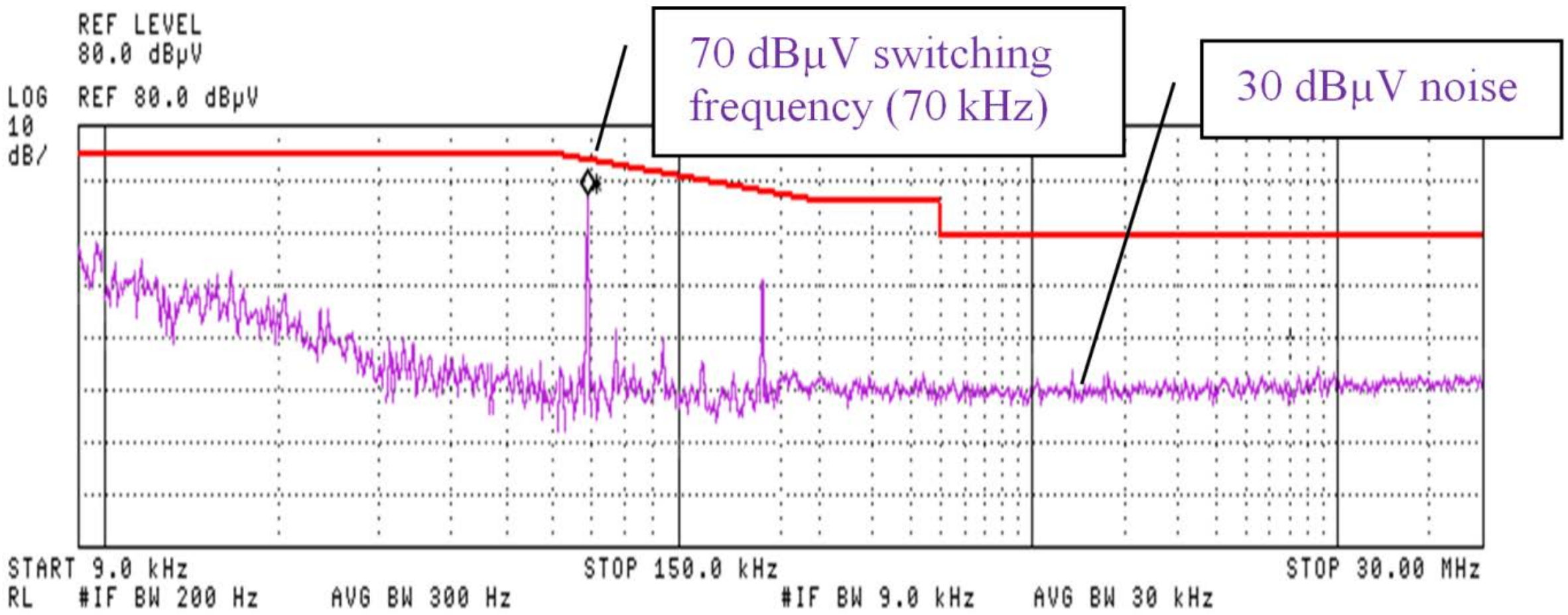

**Fig. 35:** DC output side EMC curve (1500 Ω probe)

## 7 Additional topics

This kind of converter type project is generally, and in CERN's case, put in place for providing several magnets for a given physics machine, with the additional complexity of load integrity and safety. Converters tests can then be put into perspective. The following chapters treat these topics to help the designers of converters not miss these important points. If taken into consideration at project start-up, it is possible to save a lot of time, with a real possibility of improving global project efficiency.

### 7.1 Load protection

Load protection is always a critical point, which cannot be neglected. Indeed, the high level of energy that can be stored in an inductor requires that a current path is always given to it, to avoid possible damage to the magnet (an inductive path, being opened, results in high voltage stress and dielectric damage).

CERN uses intensively the system called a crowbar, being a set of bidirectional controlled power switches in series with a resistor (though none in specific cases); this system is placed in parallel with the output of the converter, and has the function of protecting the load from a wrong condition, at the level of the power converter. A tentative opening of the load current path by the power converter would typically activate the crowbar system to avoid damaging the load. This system could be seen as replacing free-wheeling diodes in the case of a one-quadrant converter powering an inductive load. Thyristors are often chosen, since once set they are conductive (overvoltage detection across the crowbar system triggers the switches), and they will remain conductive as long as load current is present. Nevertheless, some systems use power MOSFETs as well, with the need to ensure that they are always safely controlled, letting them conduct the magnet's current. A simplified model is presented in Fig. 36.

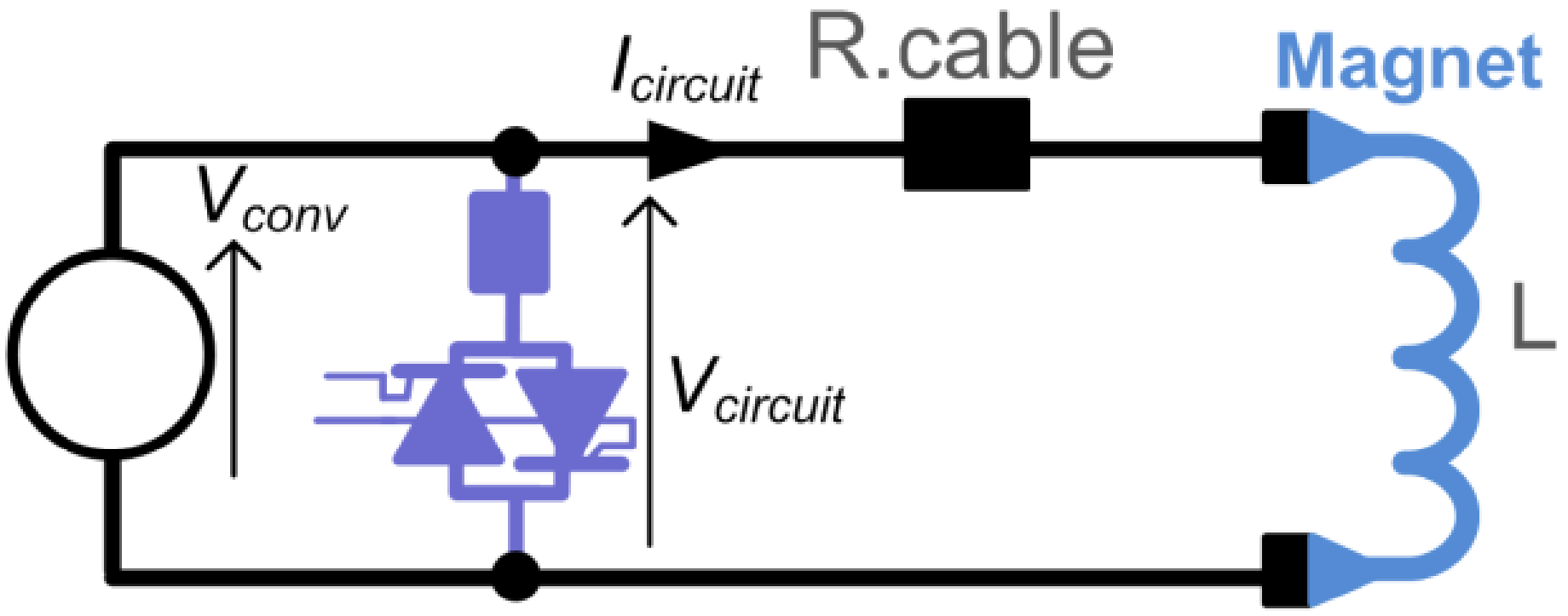

**Fig. 36:** Simplified overview of the crowbar

The design of such system, even if not complex, must ensure safe operation of the components being used. Very often, a capacitor will be placed in parallel with the crowbar system, limiting d$V$/d$t$ across it so that the speed of closing the switch, in series with the crowbar resistor, is compatible with the voltage reached across the crowbar, the power module, and finally the circuit.

In the LHC machine, power converters are modular systems that can be removed from their slots, with the risk of disconnecting a power module for replacing it (case of its failure), while a lot of current is still flowing through it. Taking such a case into account, the crowbar system was designed not to rely on the output stage power capacitors of the converter, providing its own; on the other hand, crowbar systems are sometimes placed some metres from the power converter so that they do not create conditions for oscillations (self-inductance path through the crowbar capacitors in series), and then have to be limited in their value.

In addition, the capacitors placed on the crowbar side should not lead to too high d$I$/d$t$ conditions for the thyristors when triggered, since collecting the capacitor peak current (sudden discharge) added to the long time constant's inductor current.

Figure 37 shows the capacitors (red blocks on the right-hand side) being placed away from the thyristors (left-hand side), to provide sufficient inductance in series in their path, to limit thyristor d$I$/d$t$ when turning on.

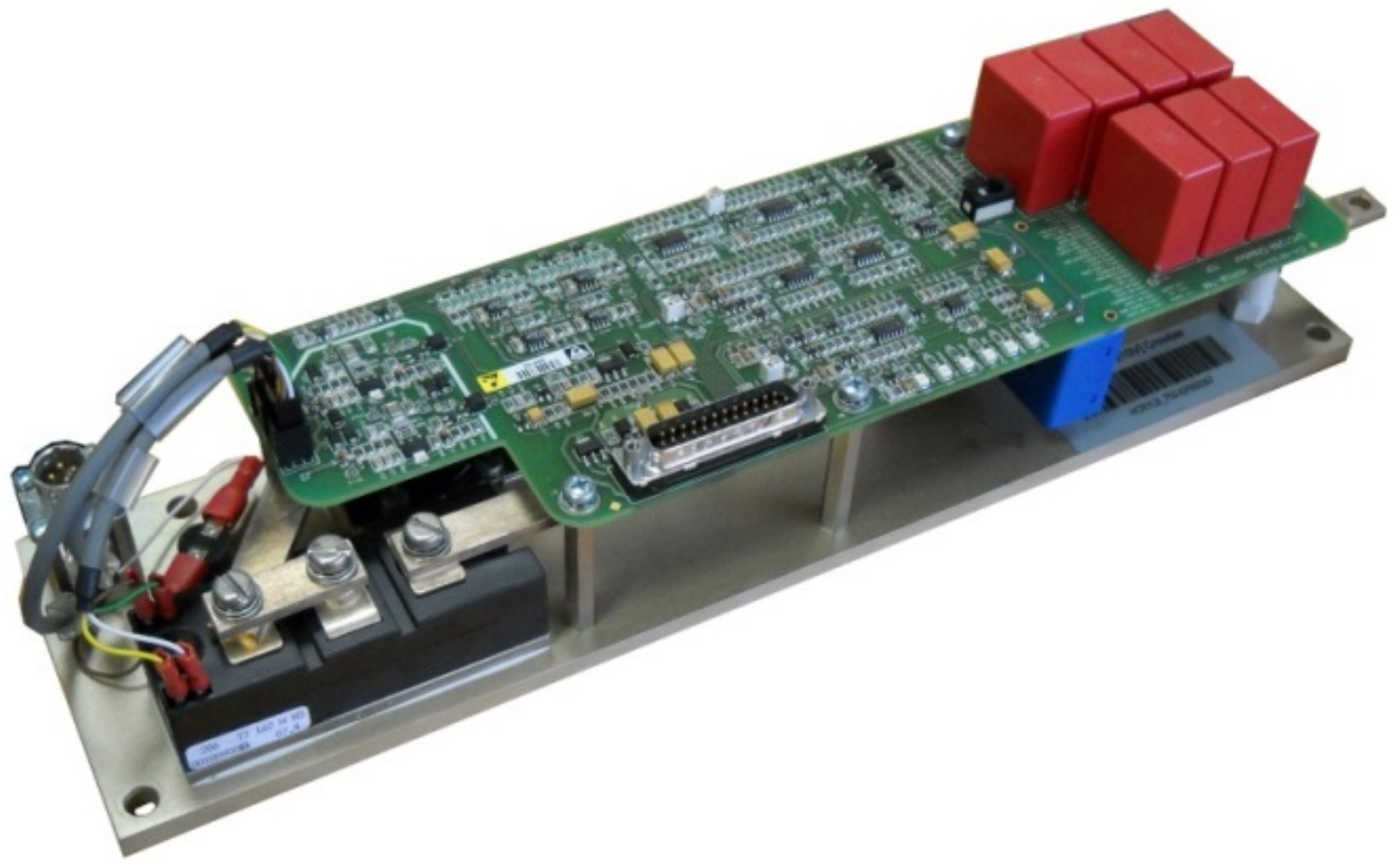
**Fig. 37:** Example of a CERN crowbar system

It can happen that power resistors can be requested to accelerate current discharge and the load energy removal process, taking into account their load energy capacity, rather than their power rating. Since the phenomena observed deal with energy versus time rather than power requirements, the function versus time of thermal management (resistor and semiconductor baseplate), and power losses inside each component (the level decreasing with time as the current is reduced), they can be considered to be unusual by designers of switching power supplies, who are more used to ‘permanent conditions’ system. A safe design must deal with these considerations, to ensure a proper level of semiconductor junction or case temperatures during the full discharge of a superconducting magnet.

## 7.2 Reception, qualification and tests

For several projects being conducted at CERN for the LHC machine, four-quadrant converters designs were validated; and tests on the series were performed with very restricted access to the final superconducting loads. CERN used some tricks to test these four-quadrant units on standard loads.

### 7.2.1 *Back-to-back test*

A four-quadrant converter is able to deliver or receive energy from an inductive load, but also from another identical four-quadrant converter. It becomes then possible to test each unit versus another unit, and reach an unlimited time constant load, with the addition of an additional control loop for one of the module assimilated to the load, and controlled as such. This additional control loop ensures that the current and voltage developed across the module are following and simulating a high time constant load, copying a superconducting magnet (Fig. 38).

If this method is very elegant and very powerful for some heat tests, to determine whether the four-quadrant part (where energy is managed or dissipated) is correctly designed, the control of inherent converter and the one added for simulating the superconducting load must be very robust, since it is probable that any instability initiated by one of the power modules would be caught by the other one, which would be of identical design, with the same weaknesses, if any. It appears that a decoupling inductor is generally required (self-inductance of cables can be sufficient), its value being determined by the bandwidth of the additional simulating load controller.

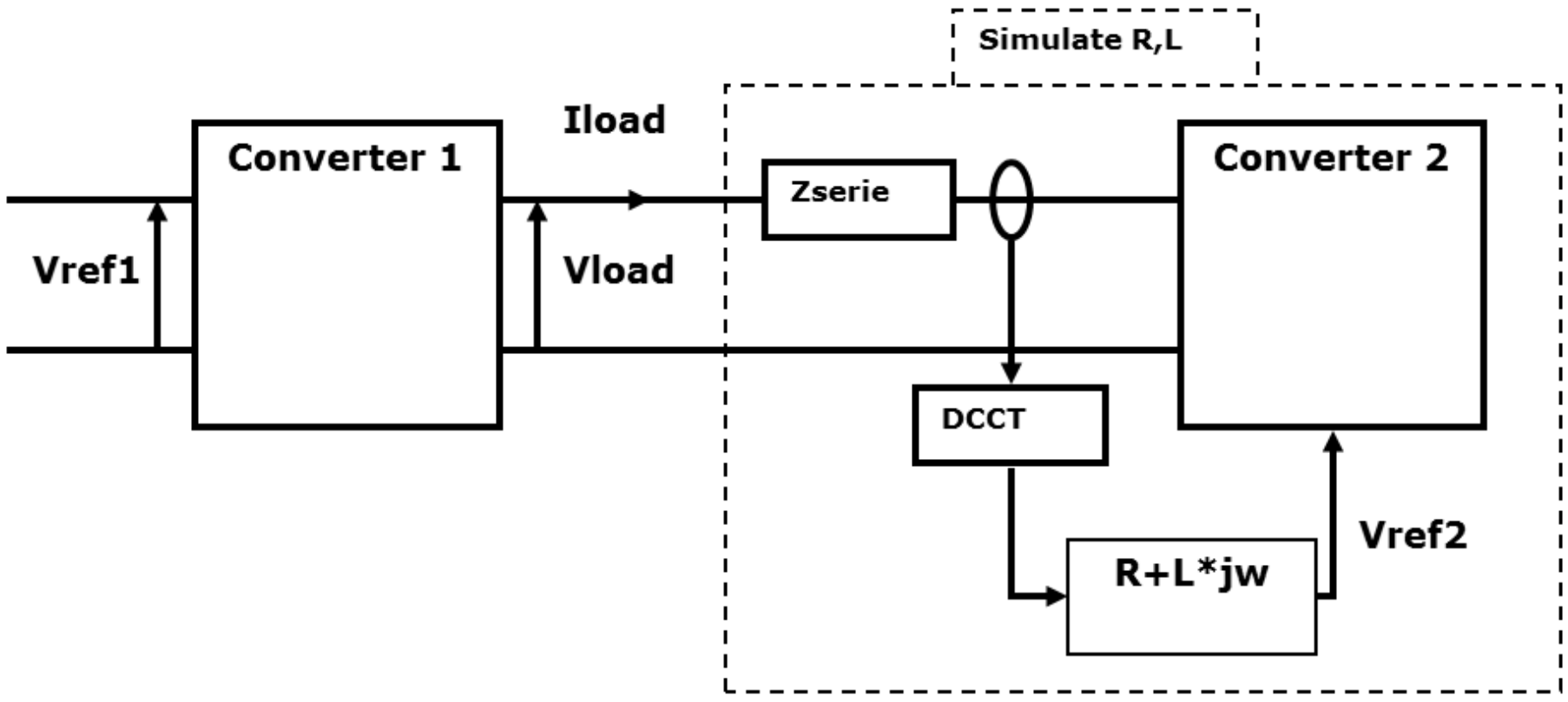

**Fig. 38:** Two converters operating back-to-back

### 7.2.2 *Four-quadrant converter bandwidth characterization*

If determining the voltage bandwidth of a converter is in general a relatively easy task, determining the behaviour of a four-quadrant converter becomes a more complex task; indeed, the converter should demonstrate that it regulates its output voltage in generator or receptor mode, leading to test facility constraints. Ideally, a converter should be operating in a receptor quadrant, in a steady state, while receiving input stimulus on its reference, to allow a Bode diagram to be plotted. This can lead to some potential issues regarding the fact that a power module is generally not designed to stay powered at a DC receptor point, the design being optimized based on the fact that load energy will decrease. That having been said, the total amount of energy that the converter can handle (except if sent back to the mains) is usually limited (by design), and in consequence, the time to measure the performance of the converter is also limited.

A simplified method consists of analysing a step-response to deduce the bandwidth of the converter alone. This simple test, even if not as rich as a frequency systematic Bode diagram type test, is often sufficient to obtain the stability margin and main performances (speed) of a system.

Taking this into account it is possible to reach the high dissipative area [V; I] of a four-quadrant converter on a standard inductive load (a few milliHenry plus a few milliOhms), with the condition of a large voltage step. A method has been put in place with two signal generators in series, providing a large step voltage (to select the receptive quadrant) superposed with a small signal square waveform (to study the step-response) (Fig. 39). The method allows the obtaining of any points for the receptive quadrant, with the combination of the current before the large step occurs, and the value of this step, in a very efficient way.

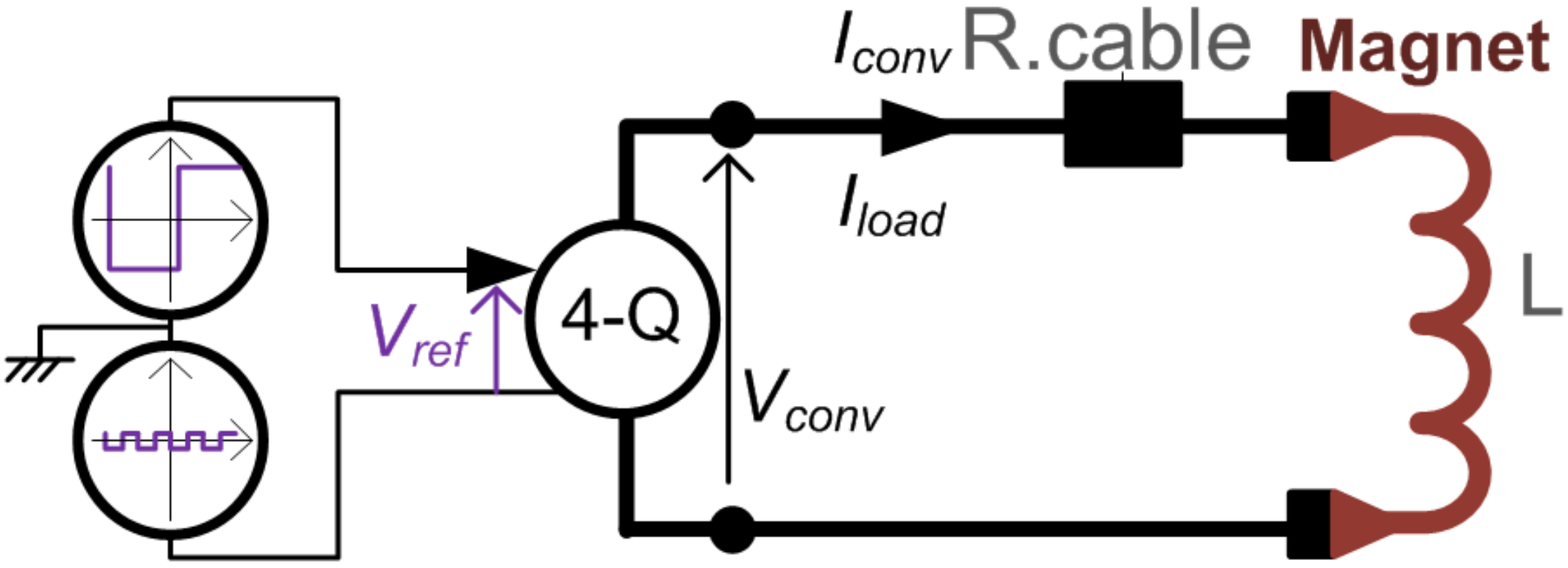

**Fig. 39:** Characterization of voltage loop using two signal generators

The result is the sinusoidal and square small signals, which are shown in Fig. 40. These were obtained from a superconducting magnet, but would have been equivalent to that obtained from a warm magnet of just 3 mH, which would be available in a standard test laboratory.

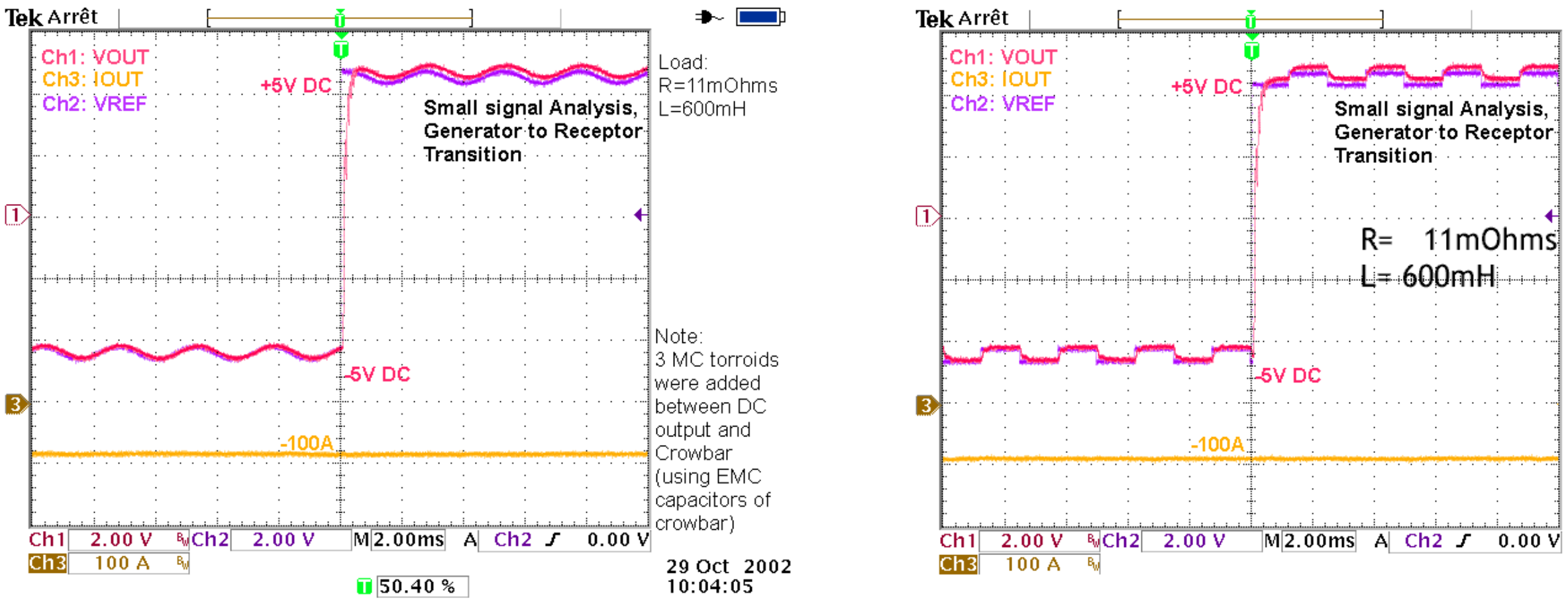

(a) (b)

**Fig. 40:** Example of step-response on a high time constant load, with the generator to receptor quadrant transition (a) on sinus type small signal; (b) on square type small signal.

## 8 Perspectives

CERN started a new project in 2013 for advanced four-quadrant power converters, using the same principles as those one described in this paper, but with an additional redundancy feature. It is intended that this will capitalize on the four-quadrant power stage using power MOSFETs and controlled as a pure current source, which it is by nature, and no longer a programmable resistor. Achieving this will allow a current source to be naturally placed in parallel, with the capability of a redundant four-quadrant power converter.

## 9 Conclusions

This paper briefly described four-quadrant power converter topologies. Critical technical points from a CERN internal design were presented, with practical results obtained on a series of converters produced for the LHC accelerator. The requirement for not having any disturbances in the operating zone (especially around [0 A; 0 V]) was achieved, thanks mainly to the circulating current loop, in its function to avoid dead zones, and to optimize MOSFET control.

The design presented was one of the less noisy converters in the LHC, which was, after some years of operation, considered to be a wise approach. EMC compliance was found to be a very important point, for example while looking for explanations of beam instability, while tracking every system regarding its potential to disturb the beam at specific medium frequencies, in the range 1–50 kHz..